\documentclass{nature2}

\usepackage{graphicx}

\usepackage{bm}
\usepackage{amsmath}
\usepackage{amssymb}
\usepackage[latin1]{inputenc}
\usepackage{color}
\usepackage[colorlinks=true, pdfstartview=FitV, linkcolor=blue, citecolor=blue, urlcolor=blue]{hyperref} 

\date{\today}

\title{Topological domain walls in helimagnets}

\author{P. Schoenherr,$^{1}$ J. M\"uller,$^{2}$ L. K\"ohler,$^{3}$ A. Rosch,$^{2}$ N. Kanazawa,$^{4}$ Y. Tokura,$^{4,5}$ M. Garst,$^{2,3\ast}$ D. Meier$^{1,6\ast}$}

\begin{document}

\maketitle

\begin{affiliations}
\item Department of Materials, ETH Zurich, Vladimir-Prelog-Weg 4, 8093 Z\"urich, Switzerland.
\item Institut f\"ur Theoretische Physik, Universit\"at zu K\"oln, Z\"ulpicher Str. 77, 50937 K\"oln, Germany.
\item Institut f\"ur Theoretische Physik, Technische Universit\"at Dresden, Zellescher Weg 17, 01062 Dresden, Germany.
\item Department of Applied Physics, University of Tokyo, Tokyo 113-8656, Japan.
\item RIKEN Center for Emergent Matter Science (CEMS), Wako 351-0198, Japan.
\item Department of Materials Science and Engineering, Norwegian University of Science and Technology, Sem Sælandsvei 12, 7034 Trondheim, Norway.
\end{affiliations}

\begin{abstract}
$^\ast$markus.garst@tu-dresden.de; dennis.meier@ntnu.no\\
\\
A magnetic helix arises in chiral magnets with a wavelength set by the spin-orbit coupling. We show that the helimagnetic order is a nanoscale analog to liquid crystals, exhibiting topological structures and domain walls that are distinctly different from classical magnets. Using magnetic force microscopy and micromagnetic simulations, we demonstrate that - similar to cholesteric liquid crystals - three fundamental types of domain walls are realized in the helimagnet FeGe. We reveal the micromagnetic wall structure and show that they can carry a finite skyrmion charge, permitting coupling to spin currents and contributions to a topological Hall effect. Our study establishes a new class of magnetic nano-objects with non-trivial topology, opening the door to innovative device concepts based on helimagnetic domain walls.
\end{abstract}

\newpage

Lamellar phases with periodically assembled layers are found in diverse systems, having profound implications in biology, physics, chemistry, and technology~\cite{deGennes&Prost,Kleman&Lavrentovich}. The lamellar texture imparts a unique combination of order and mobility, which promotes self-organization and structure formation in living systems and polymer compounds~\cite{Nakata07}. Furthermore, it finds application in liquid-crystal displays and electro-optical devices. Order and mobility are strongly affected by the presence of topological defects arising from imperfections in the planar configuration of lamellar phases. The formation of disclinations in liquid crystals and their impact on rheological properties is a well-known example~\cite{Nielsen94}. The mesoscopic length scales of liquid crystals allows direct observation of such defects and it was shown that they can exist as individual objects or building blocks of complex networks such as grain boundaries~\cite{Bouligand1973,Bouligand1983}. 

Here we demonstrate that chiral magnets are a striking analog: they similarly possess lamellar phases and ordered topological defects. Crucially, however, the topological structures exist at the nanoscale with additional functionality arising from the magnetic order. While skyrmions --- a type of whirl in magnetization --- have been reported extensively due to their potential in next-generation nanoelectronics~\cite{Muehlbauer2009,Neubauer2009,Yu2010,Yu2011,Heinze2011,
Seki2012,Fert2013,Nagaosa2013,Wiesendanger2016}, they are only part of a larger zoo of topological defects in chiral magnets, similar to the range of defects observed in mesoscopic lamellar structures. 
In particular, we show that domain walls of helimagnetic order, first discussed in ref.~\citen{Li2012}, can comprise a well-defined chain of topological defects with a distinct skyrmion charge. 
The skyrmion charge implies an efficient coupling to spin currents and contributions to the topological Hall effect~\cite{Neubauer2009,Schulz2012,Iwasaki2014,Schuette2014,Mochizuki2014}, revealing new possibilities for domain-wall-based spintronic applications.

In chiral magnets the Dzyalo\-shinskii-Moriya interaction (DMI) twists the magnetization and leads to a helimagnetic ground state. The magnetic moments periodically wind around a certain axis, ${\bf Q}$, with the wavelength $\lambda$ determined by the DMI~\cite{Bak1980}. The overall magnetization can be visualized as a stack of equidistant sheets with uniformly oriented moments, where the orientation rotates from one sheet to the next. The morphology of the spin texture is thus analogous to the chiral lamellar structure of cholesteric liquid crystals as illustrated in Fig.~1a,b~\cite{deGennes&Prost}. The wavelength $\lambda$ associated with the spin system, however, is up to three orders of magnitude smaller than in conventional liquid crystals.

Despite this difference in size, one may expect the same types of defects to form. Smooth spatial variations of the helix axis ${\bf Q}$ simply result in a curvature of the lamellar spin structure \cite{Uchida2008}. More pronounced variations, however, may break the periodicity and induce vortices, i.e., disclinations (Fig.~1c,d), as observed in cholesteric liquid crystals. The strength of such vortices is usually parametrized by the winding angle of the helix axis on a path encircling the vortex core. As the helix axis is a director (${\bf Q} = -{\bf Q}$), half integer vortices are possible with $+ \pi$ and $- \pi$ rotations (Fig.~1c,d). Furthermore, disclinations can pair up and form edge dislocations (Fig.~1e), with the Burgers vector $B$ quantifying the distance $D$ between them.

We study the emergence and coupling of such defects within helimagnetic domain walls in the near-room-temperature chiral magnet FeGe ($T_{\rm N} = 278$~K) using magnetic force microscopy (MFM). MFM is sensitive to the out-of-plane component of the magnetic stray field associated with the helimagnetism in FeGe~\cite{Dussaux2016}. Thus, the helimagnetic order manifests as a stripe-like pattern of bright and dark lines in the MFM scans in Fig.~2a-c. The periodicity of the pattern is $\lambda =70$~nm, matching with previous bulk data~\cite{Lebech1989}. However, we find equivalent values of $\lambda$ on (100) and (111)-oriented surfaces, which indicates that ${\bf Q}$ preferentially lies within the surface plane. This orientation is different from the bulk, where ${\bf Q}$ orients along the $\langle 100 \rangle$ direction. The MFM data thus reflects a surface-anchoring of ${\bf Q}$ implying a surface reconstruction of magnetic order~\cite{Rybakov2016}. Furthermore, the measured statistical orientational distribution of ${\bf Q}$ is almost flat with a slight tendency to point along a $\langle 100 \rangle$ direction (if present) within the surface plane.

Most interestingly, the MFM scans reveal helimagnetic domain walls with a complex, but well-defined inner structure. Their structure crucially depends on the angle $\angle({{\bf Q}}_1, {\bf Q}_2)$ enclosed between the helix axes of the adjacent domains as summarized in Fig.~2. Based on our data we distinguish curvature walls (type I, Fig.~2a), zig-zag disclination walls (type II, Fig.~2b), and dislocation walls (type III, Fig.~2c). 

Walls of type I occur for small angles $\angle({\bf Q}_1, {\bf Q}_2)$ and exhibit a smooth, continuous rotation of the helix axis from ${\bf Q}_1$ to ${\bf Q}_2$ as reflected by the curved dark and bright lines in Fig.~2a. The curvature wall is clearly visible due to its brighter contrast, which indicates an enhanced magnetic stray field induced by the enforced deformation of the spin helix. Type II walls display a characteristic zig-zag pattern of alternating $\pm \pi$ disclinations. They arise for intermediate $\angle({\bf Q}_1, {\bf Q}_2)$ (Fig.~2b). At the wall the magnetic stray field is enhanced and associated distortions of the helimagnetic spin structure extend over micrometer-sized distances away from the wall. Type III walls form for large $\angle({\bf Q}_1, {\bf Q}_2)$, involving a rather abrupt transformation from ${\bf Q}_1$ to ${\bf Q}_2$ (Fig.~2c). On a closer inspection, type III walls can be identified as a chain of magnetic edge dislocations.

The length of the domain walls can reach several micrometers, exceeding the characteristic length scale of the helical structure, $\lambda$, by two orders of magnitude. Individual walls can also change their type and thereby adapt to local variations of $\angle({\bf Q}_1, {\bf Q}_2)$. An example is seen in Fig.~2c, where a dislocation wall (type III) turns into a curvature wall (type I).

The observation of the three types of domain walls strikingly corroborates the analogy between topological defects in chiral magnets and cholesteric liquid crystals~\cite{Bouligand1973,Bouligand1983}. This universality emphasizes that the domain wall formation is governed by the inherent topology arising from the lamellar structure, being independent of the involved length scales and microscopic properties.

Figure 2d presents a quantitative analysis of more than 90 measured domain walls. The larger angle $\alpha$ enclosed by the domain wall and one of the helix axes ${\bf Q}_i$ is plotted as function of $\angle({\bf Q}_1, {\bf Q}_2)$ (see Fig.~2a-c). The data shows that, at the surface, ${\bf Q}$ is oriented in all directions, resulting in a broad spectrum of angles ($0^{\circ} \leq\angle({\bf Q}_1, {\bf Q}_2)\leq 180^{\circ}$). Type I walls dominate for angles $ \lesssim 85^{\circ}$ (red dots), type II walls are realized approximately between $85^{\circ}$ and $140^{\circ}$ (blue dots) and type III walls for angles $\gtrsim 140^{\circ}$ (green dots).
Figure 2d reflects that structure and orientation of the walls are interlinked. For walls of type I and III, the angle $\alpha$ follows the bisecting line ($\alpha = \frac{1}{2} \angle({\bf Q}_1, {\bf Q}_2)$). In contrast, type II walls have a tendency to orient parallel to one of the ${\bf Q}_i$, so that $\alpha = 90^{\circ}$. Close to the transition regions around $85^{\circ}$ and $140^{\circ}$, we occasionally observe a special case of type II walls. This subgroup of walls exhibits a minimal distance $D = \lambda / 2$ between $\pm \pi$ disclinations (light-blue dots) and was discussed previously in Ref.~\citen{Li2012}. 

To understand the relation between the orientation of a wall and its topological magnetic structure, we perform 2D micromagnetic simulations (see also Supplementary Information). We determine the energy density for various domain walls as function of $\angle({\bf Q}_1, {\bf Q}_2)$ on a finite size system extending up to a cut-off distance $L=12\lambda$ away from the wall (Fig.~3).
The calculations reveal that bisecting type I walls (Fig.~3a) are lowest in energy for small angles $\angle({\bf Q}_1, {\bf Q}_2)$ (red line in Fig.~3). Type II walls with $\alpha = 90^{\circ}$ (Fig.~3b) become energetically favorable as $\angle({\bf Q}_1, {\bf Q}_2)$ exceeds $85^{\circ}$, which is in excellent agreement with the experimental data (Fig.~2d). Blue lines in Fig.~3 correspond to the energies of type II walls with varying distance $D$ between $\pm \pi$ disclinations. First, walls with $D=9\lambda / 2$ are most stable. 
As $\angle({\bf Q}_1, {\bf Q}_2)$ increases type II walls with smaller $D$ become energetically preferred. Type II walls with minimal defect distance ($D = \lambda / 2$, Fig.~3c), however, are always more costly than other configurations, which explains their rareness in our MFM data and the fact that they predominantly occur in transition regions (light blue dots in Fig.~2d).

At around $145^{\circ}$ walls of type III become energetically less costly than type II walls ($D=3\lambda / 2$). Different micromagnetic wall structures are possible, all with similar energy (Fig.~3d-g). Within our numerical accuracy, a chain of serially aligned edge dislocations, first with Burgers vector $B=3\lambda$ (see~\citen{suppl}) and then with $B=2\lambda$ (Fig.~3d), is lowest in energy. Towards larger angles, type III walls with $B=2\lambda$ and $B=\lambda$ (Fig.~3e) become almost degenerate. In addition, more complex domain wall structures with slightly higher energy arise (Fig.~3f,g). Consistent with the small energy difference, a large variety of type III walls is observed experimentally (see also Supplementary Figure S2), and the micromagnetic structure frequently changes along the wall (see Fig.~2c).  

Aside from the intriguing topological nanostructure of the domain walls, their magnetism enables emergent electrodynamics ---  a property that is not available in other lamellar structures. A topological magnetic texture is identified via the skyrmion charge density $\rho_{\rm top} = \frac{1}{4\pi} {\bf M} (\partial_x {\bf M} \times \partial_y \bf M)$ for the unit vector field of the magnetization $\bf M$ defined within the $(x,y)$ plane \cite{Nagaosa2013}.

A single skyrmion with the magnetization pointing downwards at its centre and a skyrmion charge $W= \int dx dy \rho_{\rm top}=-1$ is depicted in Fig.~4a. The same charge is obtained if the structure is embedded in a topologically trivial helimagnetic background as illustrated in Fig.~4b. This embedded skyrmion however is equivalent to a pair of edge dislocations with Burgers vector $B=\lambda$ (Fig.~4c,d). The dislocation with $B=\lambda$ can thus be interpreted as a single meron, i.e., a half-skyrmion, which carries charge $W = -\frac{1}{2}$. The sign reflects the orientation of the magnetization at the meron core. Straightforward generalization of this argument (see Supplementary Information for details) yields a general relation between the skyrmion charge of a dislocation and its Burgers vector 
\begin{align}
W_{\rm disloc.} = \frac{s}{2} {\rm mod}_2\Big(\frac{B}{\lambda}\Big)
\end{align}
where ${\rm mod}_2$ is the modulo 2 operation, and the sign, $s = \pm 1$, is determined by the orientation of $\bf M$ at the dislocation center.

In order to obtain the topological charge for domain walls, one has to add the skyrmion charges of dislocations or, alternatively, pairs of $\pm\pi$ disclinations contained within the wall. This leads to the conclusion that type I and type III walls with continuous stripes have zero skyrmion charge (Fig.~3a,d,e). The type III walls of Fig.~3f,g with broken stripes, however, have a finite charge $W$. Similarly, type II can have a finite or zero charge $W$. An odd dimensionless distance $2D/\lambda$ between $\pm \pi$ disclinations leads to $W \neq 0$ and an even distance to $W = 0$.

We predict that helimagnetic walls with a finite skyrmion charge $W \neq 0$ give rise to emergent electrodynamics for electrons and magnons, and contribute to a topological Hall effect~\cite{Neubauer2009,Schulz2012,Iwasaki2014,Schuette2014,Mochizuki2014}. Their building blocks, the magnetic disclinations and dislocations, are further expected to possess more sluggish dynamics than the usual magnetic time scales. Thus, they are expected to play a key role for the material`s ac susceptibility and magnetic relaxation dynamics~\cite{Dussaux2016}. Moreover, they might be crucial for melting processes~\cite{Janoschek2013,Wilhelm2011} and, in particular, for the sought-after explanation of the non-Fermi liquid behavior in MnSi \cite{Pfleiderer2001,Ritz2013} and FeGe \cite{Pedrazzini2007} after suppressing helimagnetism with pressure.

Our results establish a striking analogy between topological domain walls in chiral magnets and defect networks in mesoscopic liquid crystals. The walls naturally arise due to imperfections in the lamellar-like magnetic ground state with profound implications for the macroscopic properties. Enabled by their unusual magnetic structure, the walls can carry a finite skyrmion charge and, hence, efficiently interact with spin currents. The observations apply to chiral magnets in general and reveal a zoo of magnetic nano-objects with non-trivial topology -- beyond skyrmions~\cite{Nagaosa2013,Fert2013,Wiesendanger2016} -- opening the door to innovative device concepts based on helimagnetic domain walls.


\newpage
\begin{flushleft}
\textbf{Acknowledgement}
We thank M. Fiebig for direct financial support and H. Simons for discussions. The work at ETH was supported by the Swiss National Science Foundation through grants 200021-149192, 200021-137520. L.K. and M.G. were supported by SFB 1143 ''Correlated Magnetism: From Frustration To Topology''. J.M. and A.R. were supported by SFB 1238 (Kontrolle und Dynamik von Quantenmaterialien). J.M. also acknowledges support from the Deutsche Telekom Stiftung and Bonn-Cologne Graduate School of Physics and Astronomy BCGS. N.K. acknowledges funding through the JSPS Grant-in-Aids for Scientific Research (S) No. 24224009 and for Young Scientists (Start-up) No. 26886005.  

\textbf{Author contributions} P.S. conducted the MFM experiments supervised by D.M.; J.M.performed the 2D micromagnetic simulations supervised by M.G. and A.R.; N.K. grew the FeGe single crystals under supervision of Y.T; P.S., J.M., L.K., M.G. and D.M. classified the domain walls; M.G. and L.K. defined the skyrmion charge of edge dislocations; P.S., J.M., L.K., M.G. and D.M. wrote the paper; M.G. and D.M. supervised the work.  All authors discussed the results and contributed to their interpretation.

\end{flushleft}

\newpage


\begin{figure}
\centering
\includegraphics[width=0.9\textwidth]{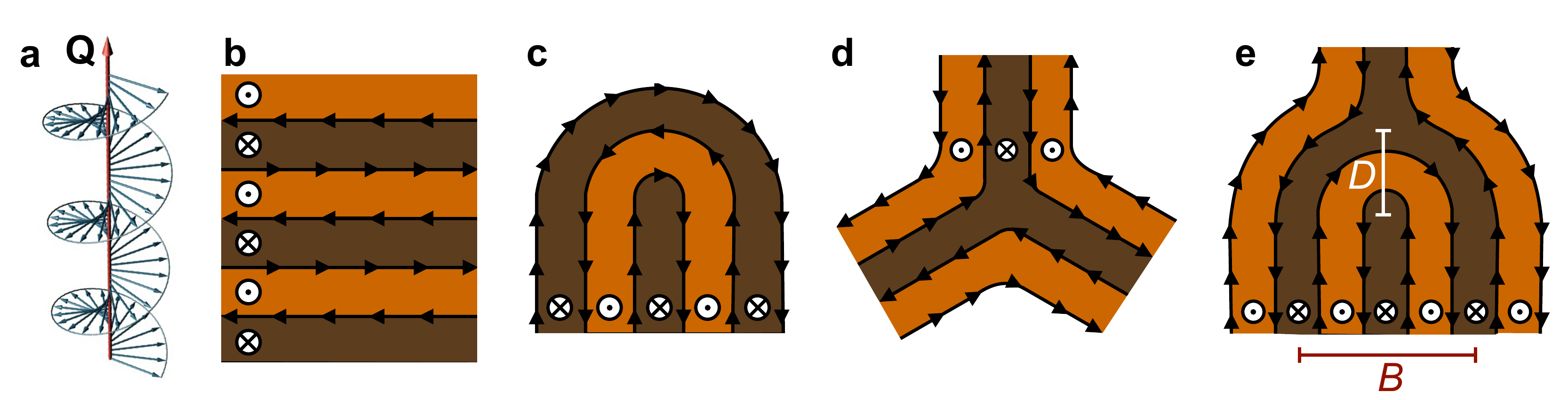}
\caption{\label{fig1}\\ {\bf{Figure 1 - Helimagnetic order and defect structures.}}
{\bf a,} Right-handed magnetic helix.
{\bf b,} Magnetic moments form periodic layers (black arrows) orthogonal to the helix axis ${\bf Q}$. Bright and dark areas corresponds to stray fields pointing in and out of the plane, respectively. {\bf c,} $\pi$ disclination. {\bf d,} $-\pi$ disclination. {\bf e,} Edge dislocation formed by a pair of $\pi$ and $-\pi$ disclinations at distance $D$. The Burgers vector, $B = 2 D$, is given by an integer multiple of the helix wavelength, $B = n \lambda$ (here, $n=2$).
} 
\end{figure}

\newpage

\begin{figure}
\centering
\includegraphics[width=0.4\textwidth]{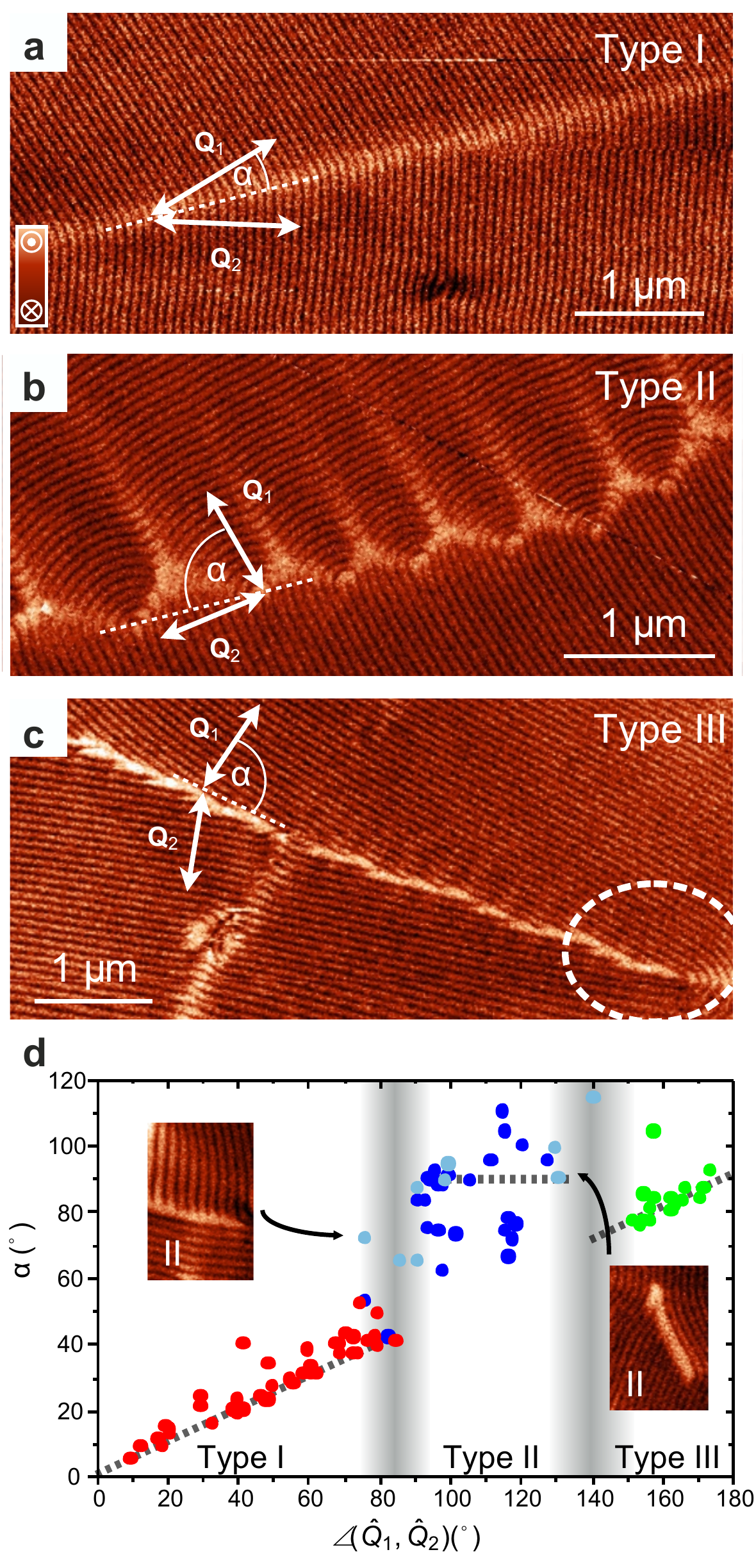}
\caption{\label{fig2}\\ {\bf{Figure 2 - Helimagnetic domain walls in FeGe.}}
{\bf a,} Curvature wall (type I). {\bf b,} Zig-zag disclination wall (type II). {\bf c,} Dislocation wall (type III). The MFM data is obtained at 260-273~K. {\bf d,} Quantitative analysis of the domain wall angle $\alpha$ (see {\bf a}-{\bf c}) as function of the angle between ${\bf Q}_1$ and ${\bf Q}_2$ ($\angle({\bf Q}_1, {\bf Q}_2)$). The plot reveals three distinct stability regimes for type I (red), type II (blue), and type III (green) walls. Special type II walls (light-blue) occur in the transition regions (see inset images) as explained in detail in the main text. Within the marked region in {\bf c}  (white dashed contour) the wall changes from type III to type I.
}
\end{figure}

\newpage

 \begin{figure}
\centering
\includegraphics[width=0.6\textwidth]{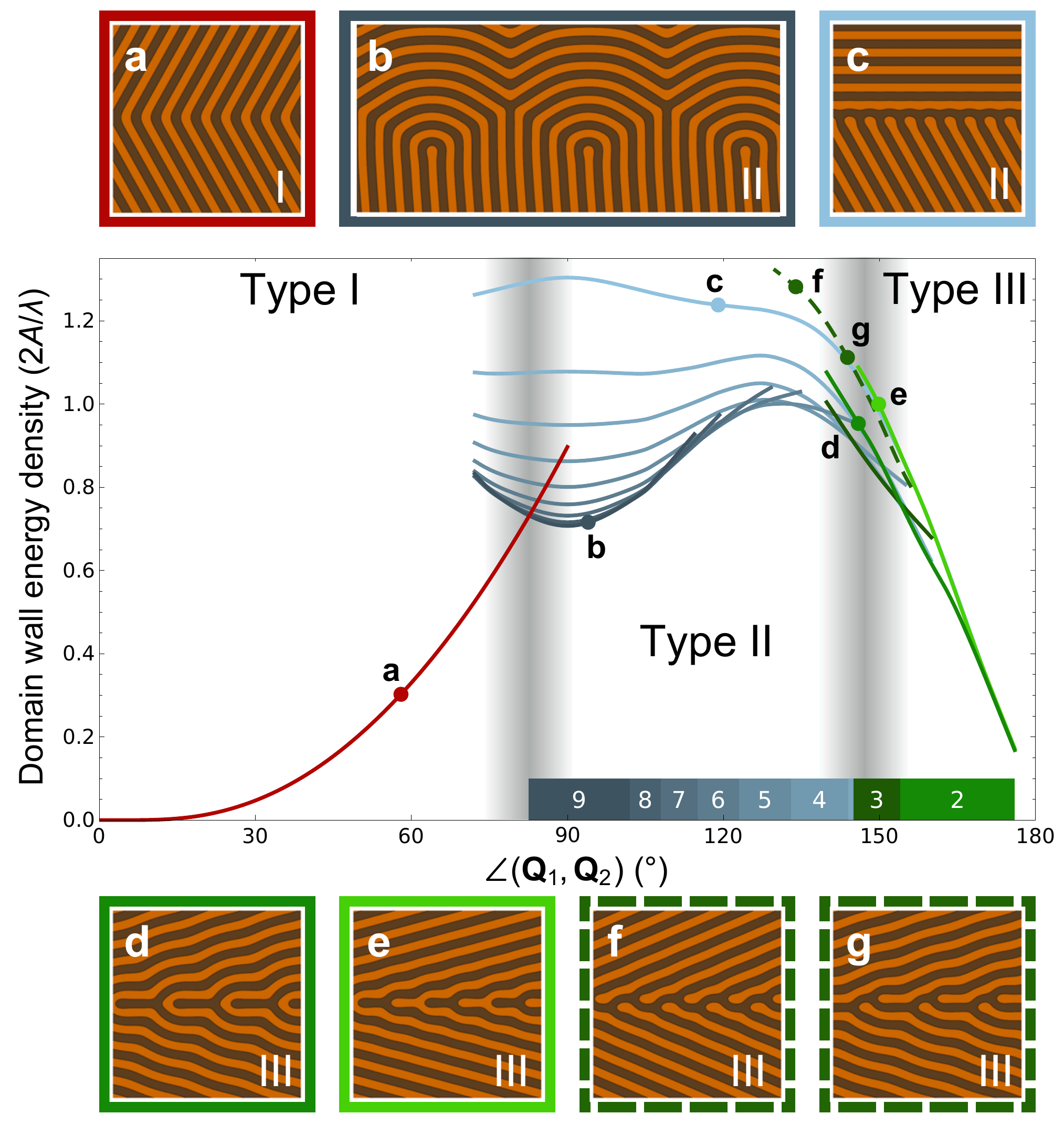}
\caption{\label{fig3}\\ {\bf{Figure 3 - Micromagnetic domain wall simulations}}
{\bf a,} type I,{\bf b,c,} type II, and {\bf c-g,} type III walls. The graph shows that type I walls are lowest in energy density for small angles $\angle({\bf Q}_1, {\bf Q}_2)$. Type II walls stabilize for intermediate angles. As $\angle({\bf Q}_1, {\bf Q}_2)$ increases, type II walls with smaller defect distance $D$ become energetically favourable (blue, numbers in units of $\lambda/2$). For larger angles, type III walls with continuous stripes ($B=3\lambda$ (dark green line), $2\lambda$ ({\bf d}) and $\lambda$ ({\bf e}) are most stable. The associated Burgers vector varies with $\angle({\bf Q}_1, {\bf Q}_2)$ (green, numbers in units of $\lambda$).
}
\end{figure}

\newpage

 \begin{figure}
\centering
\includegraphics[width=\textwidth]{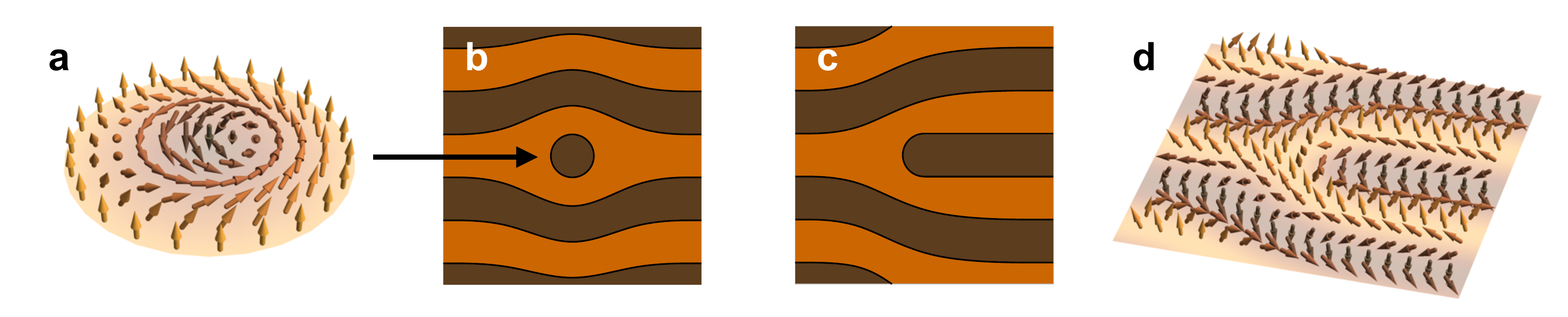}
\caption{\label{fig4} {\bf{Figure 4 - Magnetic edge dislocation with nonzero skyrmion charge.}}
{\bf a,} Magnetic skyrmion with topological charge $W=-1$, {\bf b,} Cartoon of a skyrmion embedded in a helimagnetic background, {\bf c,d} Cartoon and spin configuration of an edge dislocation with Burgers vector $B=\lambda$, i.e., a meron with $W = -\frac{1}{2}$.
}
\end{figure}


\newpage

\section*{Supplementary Text}

In this supplement, we provide details about the experimental methods and present more examples of observed domain wall configurations. The analysis of the experimental data is discussed in detail. Moreover, we describe our micromagnetic simulations and analyze the results. The distance between disclinations within type II zig-zag disclination walls is compared between theory and experiment. Finally, a detailed discussion of the topological skyrmion charge attributed to dislocations is presented, and the charge of various domain walls is specified.

\section{Experimental Methods}

\subsection{Sample preparation}

FeGe single crystals were grown by chemical vapour transport from FeGe B35 powder with I2 (20 mg) in an evacuated quartz tube. The tube was mounted in a heated three-zone furnace for 1 month with a thermal gradient of $560^{\circ}$~C to  $500^{\circ}$~C. This leads to the growth of $0.5 \times 1 \times 1$ mm$^3$ large B20 FeGe crystals at the lower temperature side. The B20 crystal structure was confirmed by Laue diffraction. For our magnetic force microscopy studies, samples with $(100)$- and $(111)$-oriented surfaces were cut and chemo-mechanically polished with silica slurry, yielding flat surfaces with roughness of $\approx$ 1 nm.

\subsection{Magnetic force microscopy - MFM}

For our MFM measurements we used magnetic coated tips with a tip radius $\lesssim 50$~nm (Nanosensors, PPP-MFMR, resolution $<50$~nm). Imaging was performed in two-pass mode. First the topography was recorded in semi-contact. In the second path the magnetic response was detected with a fixed tip-surface distance of about 30 nm. The magnetic tips were operated at their resonance frequency of 75-77 kHz and with a scan speed of $2-3.5$~$\mu$m s$^{-1}$ to achieve optimal image resolution of $10-15$~nm. All MFM images were measured in a commercially available scanning probe microscope (NT-MDT) using a home-built cooling holder. This cooling holder is based on a water-cooled three-stage peltier element and reaches temperatures down to 260~K, giving us access to the helimagnetic phase of FeGe ($T_{\rm N} = 278$~K). Low water flow rates were used in order to reduce noise from vibrations due to the water cooling cycle. Additionally, the measurements were performed in a dry nitrogen environment (humidity below 1~\%) to avoid ice formation on the sample surface.

\subsection{Examples of observed domain wall configurations}

In Fig. 2 of the main text, typical examples of the three types of domain walls are shown.
Figure S1a and c presents the same walls but on a larger scanning area ($\sim 6\ \mu$m x $6\ \mu$m). Panel b shows a different example of a type II wall, and it covers the whole length of the wall including its beginning and end.
In comparison to type I and III, the width of type II walls is relatively large. It consists of the extent of the zig-zag pattern itself as well as the $\mu$m-sized distortions emanating from the $-\pi$ disclinations affecting the regular stripe pattern of the ${\bf Q}_1$ domain in panel b.

Besides these representative domain walls, there is a wide variety of domain walls observed by our MFM measurements. Figure S2 identifies some of the subtypes for the three domain wall categories. Panel a shows a curvature wall with a large angle $\angle({\bf Q}_1, {\bf Q}_2)$ enclosed by the helix axes of the two domains that splits into two type I walls with smaller angles. In panel b a type II wall with its characteristic zig-zag pattern meets a type III dislocation wall seen in the upper right corner. Panel c and d show irregular type II domain walls. Note that there exist helimagnetic stripes that can be continuously followed from one domain to the other resulting in a deviation of the domain wall angle $\alpha$ from $90^\circ$. In panel e the domain wall angle $\alpha$ of a type III dislocation wall varies along the wall. In general, one can distinguish type III walls with continuous and discontinuous stripe patterns, see in the main text Fig. 3d,e and f,g, respectively. Whereas in Fig.~S2f and g the type III walls contain regions with continuous stripes, the dark stripes in panel h are all discontinuous.

\subsection{Statistical analysis of domain walls}
\label{sec:StatAnalysis}

\subsubsection{Domain wall angle $\alpha$}
In Fig.~2d of the main text, the relation between the domain wall angle $\alpha$ and the angle $\angle({\bf Q}_1, {\bf Q}_2)$ enclosed by the helix axes of the two domains is analysed. Each point in this diagram corresponds to a single wall. The lengths of the walls varies between $\approx 1-21$~$\mu$m. Arrow bars reflect the accuracy of the angles extracted from the MFM images, that is, better than $\pm 5^{\circ}$. For domain walls of type I and III the variance of the experimentally observed values around the bisecting line $\alpha = \frac{1}{2}\angle({\bf Q}_1, {\bf Q}_2)$ is relatively small. The type I curvature walls are free of defects so that they are expected to relax easily in order to 
assume their energetically most stable configuration. 
Similarly, the dislocations within a type III wall are able to arrange themselves by a climbing motion to yield an optimal domain wall angle\cite{Dussaux2016}, which does not involve the crossing of a topological barrier. In contrast, the variance of the data related to type II walls is comparatively large which will be discussed in detail in the following section.

\subsubsection{Statistical analysis of type II zig-zag disclination walls}

In order to evaluate the micromagnetic domain wall structure of walls of type II, we performed the following statistical analysis. Consider a type II wall oriented such that the $+\pi$ defects are located below the $-\pi$ defects, see Fig.~S3a. Then, we define for the $i^{\rm th}$ $+\pi$ disclination of the wall the integers $x_i$ and $y_i$ measuring the distance to the adjacent $-\pi$ disclination to the left- and right-hand side, respectively, in units of $\lambda/2$. The average of these values over a single domain wall are denoted by $\bar{x}$ and $\bar{y}$. 

The mean distance between adjacent disclinations within a single wall is then given by $D = \frac{\lambda}{2} \frac{\bar{x}+\bar{y}}{2}$. The dependence of $D$ on $\angle({\bf Q}_1, {\bf Q}_2)$ is discussed in Section \ref{sec:DisclinationDistances} and compared to theory, see Fig.~S9. Error bars for each domain wall are given by the standard deviation and depend on the domain wall length. The domain wall lengths used in our statistical analysis vary from 3 to 44 $\pi$ defects (average length of 14 $\pi$ defects per wall).

The distance between $\pm \pi$ disclinations fluctuates within a wall as can be seen in Fig.~S3a. In order to quantify these fluctuations, we consider in Fig.~S3b the difference $\bar{x} - \bar{y}$ for each wall. We find that the mean $\langle \bar{x} - \bar{y} \rangle \approx -0.1$ averaged over all walls is basically zero with a standard deviation $\sigma_{\bar{x} - \bar{y}} \approx 3.5$ indicated by the grey shaded region in panel b.

In the main text, we introduced the domain wall angle $\alpha$ whose definition was given in Fig.~2 of the main text. The dependence of the experimentally determined $\alpha$ as a function of $\angle({\bf Q}_1 , {\bf Q}_2)$ was discussed in Fig. 2d of the main text. The domain wall angle $\alpha$ is related to 
$\bar{x}$ and $\bar{y}$ by the formula
\begin{align}
\alpha = \frac{\pi}{2} + \arctan \Big(\frac{(\bar{x} - \bar{y}) \sin \angle({\bf Q}_1 , {\bf Q}_2)}{\bar{x} + \bar{y} +(\bar{x} - \bar{y}) \cos \angle({\bf Q}_1 , {\bf Q}_2)}\Big).
\end{align}
It follows that for a symmetric type II wall with $\bar{x} = \bar{y}$ the domain wall angle is $\alpha = 90^\circ$. On average this holds approximately as $\langle \bar{x} \rangle \approx \langle \bar{y}\rangle$. Moreover, for a zero average $\langle \bar{x} - \bar{y} \rangle = 0$ the standard deviation of $\alpha$ is simply given by 
\begin{align}
\sigma_\alpha = \frac{|\sin \angle({\bf Q}_1 , {\bf Q}_2)| }{2 d(\angle({\bf Q}_1 , {\bf Q}_2))} \sigma_{\bar{x} - \bar{y}}
\end{align}
where $d(\angle({\bf Q}_1 , {\bf Q}_2))$ is the average of $(\bar{x} + \bar{y})/2$ over domain walls at a given $\angle({\bf Q}_1 , {\bf Q}_2)$. This implies a standard deviation $\sigma_\alpha$ ranging approximately between $20^\circ$ and $38^\circ$ for angles $\angle({\bf Q}_1 , {\bf Q}_2)$ between $90^\circ$ and $130^\circ$. This is consistent with the observation in Fig.~2d of the main text.

The above analysis of the experimental data indicates that zig-zag disclination walls prefer a domain wall angle $\alpha = 90^\circ$. For this reason, we limited our micromagnetic simulations of type II walls to this value of $\alpha$. The relatively large standard deviation 
$\sigma_\alpha$ is related to the fact that the relaxation of a type II wall with $\alpha \neq 90^\circ$ requires a specific rearrangement of disclinations which is however inhibited by a topological barrier.

\section{Theoretical considerations}
\label{lit:appendix-numerics}

\subsection{Micromagnetic simulations of helimagnetic domain walls}
\label{sec:MicroSim}

For a quantitative comparison with experiment, we performed micromagnetic simulations of various domain wall structures of helimagnetic order. In FeGe the magnetic exchange is predominant followed by a 
significantly weaker Dzyaloshinskii-Moriya interaction. This leads to a variation of the magnetization on a length scale substantially larger than the atomic unit cell. This allows for a continuum description in terms of a continuous vector field ${\bf n}(x,y)$ of unit length $|{\bf n}|=1$ representing the orientation of the magnetization vector. From the experiments we know that the pitch of the helix is aligned parallel to the surface so that we restrict ourselves, for simplicity, to an effective two-dimensional simulation within the surface plane. We consider a minimal model in terms of the free energy functional
\begin{equation} \label{Theory}
 F[{\bf n}] = \int\!\mathrm{d}^2{\bf r} \Big(A (\partial_\alpha {\bf n}_i)^2 + \mathcal{D} \, \varepsilon_{i \alpha j} {\bf n}_i \partial_\alpha {\bf n}_j\Big)
\end{equation}
where indices $i,j=x,y,z$ sum over all components of the magnetization while the spatial index $\alpha=x,y$ is confined to the two-dimensional plane. $\varepsilon_{i \alpha j}$ is the totally antisymmetric tensor with $\varepsilon_{xyz} = 1$. $A$ is the exchange stiffness and $\mathcal{D}$ is the Dzyaloshinskii-Moriya interaction. For simplicity, we neglect dipolar interactions and magnetic anisotropies. 

The helimagnetic ground state of Eq.~\eqref{Theory} is given by ${\bf n}({\bf r}) = {\bf e}_1 \cos(Q {\bf e}_3{\bf r}+\varphi) + {\bf e}_2 \sin(Q {\bf e}_3{\bf r}+\varphi)$ where ${\bf e}_3$ is lying within the plane,
the orthonormal unit vectors obey ${\bf e}_1 \times {\bf e}_2 = {\bf e}_3$, and the helix wavelength is given by $\lambda = 2\pi/Q$ with the wavevector $Q = \mathcal{D}/(2A)$. The phase $\varphi$ can be chosen arbitrarily as it is a zero mode, i.e., the energy does not depend on its choice. The helix axis ${\bf Q}$ is determined by ${\bf e}_3$. Note that the helix is invariant with respect to a 
$\pi$-rotation of the three orthonormal vectors around ${\bf e}_1$ (for $\varphi = 0$), i.e., ${\bf e}_1 \to {\bf e}_1$, ${\bf e}_2 \to -{\bf e}_2$, ${\bf e}_3 \to -{\bf e}_3$. The helix axis ${\bf Q}$ is thus only defined up to a sign so that it is effectively a director field allowing for half-integer vortices, in particular, $\pm\pi$ disclinations.

After measuring energy and length in dimensionless units, $\tilde F = F/(2A)$ and $\tilde r = r Q$, the theory \eqref{Theory} is parameter-free. This implies that the energy of a particular domain wall between two helimagnetic regions is exclusively determined by geometric factors: the relative angle between the two domains and the angle defining the orientation of the wall, denoted in the main text  by $\angle({\bf Q}_1, {\bf Q}_2)$ and $\alpha$, respectively, (see Fig. 2 of the main text).

We determine the energy of various domain wall configurations numerically using micromagnetic simulations of the Landau-Lifshitz-Gilbert equation
\begin{equation}
\partial_t  {\bf n}= - {\bf n} \times {\bf B}_{\rm eff} + \alpha_G\, {\bf n} \times \partial_t {\bf n}
\label{eq:LLG}
\end{equation}
with the dimensionless effective magnetic field ${\bf B}_{\rm eff} = - \delta \tilde F/\delta {\bf n}$, and the dimensionless time $t$. The continuous field ${\bf n}$ is discretized with the help of classical Heisenberg spins on a square lattice with lattice spacing $a \approx 0.024\lambda$. Derivatives are approximated by finite differences, and the total grid is rectangular with $N_x \times N_y$ sites.
The Gilbert damping is chosen within the range $\alpha_G \in [0.1,0.8]$.

We choose the domain wall to be aligned with the $x$-axis. Anticipating periodic domain walls, we employ periodic boundary conditions along the $x$-axis enforcing a periodicity of the wall which is commensurate with the size $N_x$. The boundary conditions along the other two edges 
are fixed and chosen such that the resulting spin configuration corresponds to magnetic helices with orientations ${\bf Q}_1$ and ${\bf Q}_2$ with certain phases $\varphi_1$ and $\varphi_2$. Moreover, the angles $\angle({\bf Q}_1, {\bf Q}_2)$ and $\alpha$ have to be restricted to specific values so that the two boundary helices are commensurate with the chosen system size. 

The interface between these two helimagnetic domains is first initialized with a magnetization profile that implements the corresponding topology of the wall. Afterwards the system evolves according to Eq.~\eqref{eq:LLG} and relaxes the implemented magnetization yielding the energy. In order to obtain the energy $\tilde E_{\rm dw}$ attributed to the domain wall, the energy of the perfectly ordered helimagnetic state is subtracted. Small systematic errors might arise here from anisotropy energies due to the discretization on a quadratic lattice. Note that the result for the domain wall energy does not depend on the choice of phases, $\varphi_1$ and $\varphi_2$, as the wall can adapt its position along the $x$- and $y$-axes.

Preferably, the distance between the domain wall and the boundaries, $L$, should be chosen large enough such that the fixed boundary configurations do not significantly influence the energetics. We observe that the finite size scaling substantially depends on the type of wall under consideration and, in particular, on the width of the wall. 
The energy converges with increasing $L$ the slower the larger the width of the domain wall.

A slow convergence is especially observed for type II walls which possess a large width for 
$\angle({\bf Q}_1,{\bf Q}_2) \approx 90^\circ$. As an example with particularly strong finite-size correction, we present in Fig.~S4 a single period of such a wall with disclinations separated by $D = 9\lambda/2$ 
after full numerical optimization for three different system sizes. Similar to Fig. 1 of the main text, we only show a binary representation of the full helimagnetic structure where bright and dark stripes represent regions where the local magnetization possesses a component that points out or into the plane, respectively. The distance $L$ between the upper fixed boundary and the domain wall interface corresponds to 12, 15, and 20 helix wavelengths $\lambda$ for panels a, b, and c, respectively. 
The deformation of the helix arrangement induced by the $-\pi$ disclination extends rather far away from the wall, which is also observed experimentally, see Fig.~S1b. This leads to a relatively strong dependence of the energy density on the system size $L$, see Fig.~S7a, amounting to differences of $14\%$ for the configurations shown in Fig.~S4. In contrast, for walls with a comparatively smaller width as shown in Fig.~S5 and S6 
the finite-size corrections are substantially smaller for the same values of $L$, see Fig.~S7b and c, with a difference in energy density of $6\%$ and $0.6\%$, respectively.
As the evaluation of domain wall energies is computationally demanding, we limit ourselves in the main text to a comparison of energies for a system size of $L = 12 \lambda$. The precise transition angles between the various walls and details of the spectrum might thus alter in the thermodynamic limit of very large systems, see also the discussion in Section \ref{sec:DisclinationDistances}. However, we expect that the qualitative features remain unaffected.

The micromagnetic simulations are nevertheless in remarkably good agreement with our experimental findings.  
This is noteworthy because our two-dimensional simulations do not only neglect dipolar interactions but also do not account for the additional twist of the magnetization on the surface of a three-dimensional bulk helimagnet which arises from the Dzyaloshinskii-Moriya interaction\cite{Wilson2013, Rohart2013}. The helix gains additional energy from such surface twists that also influence the energetics of domain wall configurations within the surface plane. In fact, we believe that the energy gain from such surface twists is responsible for the surface anchoring of the helix axes observed in our experiments on FeGe.

\subsection{Domain wall configurations}

In Fig. 3 of the main text, the numerically obtained energy density and various domain wall configurations were presented for a system size of $L = 12 \lambda$. In Fig. S8 we again show this energy spectrum and an extended set of real space configurations. Panels a to l show a series of energetically minimal domain wall structures as a function of increasing angle $\angle({\bf Q}_1, {\bf Q}_2)$. The zig-zag disclination walls in panel c to h are distinguished by a decreasing distance $D$ between $\pi$ and $-\pi$ disclinations starting from $9$ and ending with $4$ in units of $\lambda/2$. 
The extent of the various type II walls as a function of $\angle({\bf Q}_1, {\bf Q}_2)$ is indicated by the blue shaded boxes right above the $x$-axis  in Fig.~S8. Upon increasing $\angle({\bf Q}_1, {\bf Q}_2)$ further, one first observes the type III wall of panel i containing a chain of serially aligned edge dislocations with Burgers vector $B = 3 \lambda$. This is followed by a chain of dislocations with Burgers vector $B = 2\lambda$ whose separation increases with further increasing $\angle({\bf Q}_1, {\bf Q}_2)$ as indicated in panel j, k and l. Within our numerical accuracy, we find that the wall containing a chain of dislocations with $B = \lambda$ shown in panel t always has a slightly higher energy than $B = 2\lambda$ although the energies approach each other very closely for large $\angle({\bf Q}_1, {\bf Q}_2)$. 
Panels m to s  show further configurations that possess a higher energy than the domain wall ground states.

\subsection{Distance between $\pm \pi$ disclination defects within type II domain walls: theory vs. experiment}
\label{sec:DisclinationDistances}

The blue shaded boxes above the $x$-axis in Fig.~S8 indicate in which regime of angles $\angle({\bf Q}_1, {\bf Q}_2)$ type II zig-zag disclination walls with a certain distance $D$ between $\pm\pi$ disclinations are expected to be most stable. In Fig.~S9, this theoretical result for a series of special values of $\angle({\bf Q}_1, {\bf Q}_2)$ (red and green circles) is compared with $D$ obtained from a statistical analysis of experimentally observed zig-zag disclination walls (blue and light blue triangles), see Section \ref{sec:StatAnalysis}. Theoretically, only type II walls with a constant distance $D$ between defects are considered whereas in the experiment the distance between disclinations fluctuates within the wall giving rise to the error bars.
The vertically grey-shaded areas again indicate approximately the transition region towards type I and type III walls for smaller and larger values of $\angle({\bf Q}_1, {\bf Q}_2)$, respectively.
Within these transition regions type II walls with small distances $D \gtrsim \lambda/2$ are observed which are represented by the light blue triangles. Theoretically, we find however that such walls have a relatively large energy. The configurations with larger distances $D$ are found to possess a smaller energy and they are more frequently observed (blue triangles). Consistently in the experimental as well as the theoretical data, the distance $D$ is maximal for $\angle({\bf Q}_1, {\bf Q}_2) \approx 90^\circ$, and it decreases for smaller and larger values of $\angle({\bf Q}_1, {\bf Q}_2)$. The precise theoretical values of $D$ increase with increasing system sizes; the red and green dots were obtained for $L = 12\lambda$ and $L=20\lambda$, respectively, see the discussion in Section \ref{sec:MicroSim}. Domain wall configurations up to $2D/\lambda = 10$ have been simulated; the distance $D$ of the optimal configuration probably exceeds this value
for $L=20\lambda$ close to $\angle({\bf Q}_1, {\bf Q}_2) \approx 90^\circ$.  
Whereas the qualitative trend agrees between theory and experiment, it seems that the asymptotics for $L \to \infty$ quantitatively overestimates the experimentally determined values of $D$. This might hint at the presence of an intrinsic cutoff length, or it indicates that contributions neglected in our numerical simulations are important for a full quantitative agreement, see the discussion at the end of Section \ref{sec:MicroSim}.

\subsection{Skyrmion charge of a dislocation}

The topological skyrmion charge density is defined by $\rho_{\rm top} = \frac{1}{4\pi} {\bf n} (\partial_x {\bf n} \times \partial_y {\bf n})$. A finite topological charge gives rise to an emergent electrodynamics in the interplay of electron spin currents or magnon currents with a magnetic texture\cite{Neubauer2009, Schulz2012, Mochizuki2014, Jonietz2010}, see N. Nagaosa et al.\cite{Nagaosa2013} and M. Garst\cite{Garst2016} for reviews.
We can attribute a well-defined skyrmion charge to any localized magnetic texture provided that its associated charge density decays sufficiently fast with increasing distance $\rho = \sqrt{x^2 + y^2}$ so that the two-dimensional integral $\int dx dy \rho_{\rm top}$ converges.

This is for example not the case for disclinations. This is best illustrated for a $2\pi$ disclination that we can approximate with the parametrization
\begin{align} \label{2pi}
{\bf n}_{2\pi}(\vec r) = (\sin \phi \sin(\theta_0 - Q \rho), - \cos \phi \sin(\theta_0 - Q \rho), \cos(\theta_0 - Q \rho))
\end{align}
where $\vec r = \rho (\cos \phi, \sin \phi)$. We consider the cases $\theta_0 = \pi$ and $\theta_0 = 0$ for which the magnetization at the disclination core points down and up, respectively, see Fig.~S10. Its topological charge density
\begin{align}
\rho^{2\pi}_{\rm top} = - \frac{Q}{4\pi \rho} \sin(\theta_0 - Q \rho)
\end{align}
only decays as $1/\rho$, and, as a result, its topological charge is not well-defined.  This is reflected in the fact that the integral 
over the topological density depends on the integration area even for large $\rho$. The integral over the area of a circle around the origin with radius $m \pi/Q$ yields zero for even integer $m$. For odd integer $m$ the resulting topological charge is $-1$ for $\theta_0 = \pi$ and $+1$ for $\theta_0 = 0$.

The topological charge of a dislocation, however, is well-defined because its topological charge density decays faster than $1/\rho$ as the two $\pm\pi$-dislinations screen each other at long distances. 
For its determination we invoke the concept of topological, i.e., continuous transformations. First, consider the pristine helix which varies only along a single direction in space, so that its topological charge density is exactly zero. As a result, the total charge integrated over a large area also vanishes. This remains true after applying a topological transformation of the helix by elastically deforming it within a locally confined area, see Fig.~S11. Whereas the deformed configuration in Fig.~S11b will have a finite local skyrmion charge density, its total charge integrated over the full area remains zero.

Now let us distort the helix locally like in Fig.~S11b and insert a locally confined $2\pi$ disclination of radius $m \pi/Q$. This leads to magnetic textures as shown, for example,  in Fig. S12a, c, e, and g. It follows that the total topological charge of the whole magnetic texture is determined by the embedded $2\pi$ disclination only, and it is $0$ for even $m$ and $\pm 1$ for odd $m$ where the sign depends on the orientation of the magnetization at the disclination core. However, the resulting texture can be alternatively identified as a helix containing two dislocations of the type shown in panels b, d, f, and h, 
with Burgers vector $B = m 2\pi/Q = m \lambda$. By symmetry, it follows that each of them must carry half of the total charge. This yields the formula for the skyrmion charge of a dislocation given in the main text 
\begin{align} \label{ChargeDisl}
W_{\rm disloc.} = s \frac{1}{2} {\rm mod}_2 \frac{B}{\lambda}.
\end{align}
The sign $s$ is determined by the magnetization at the associated $2\pi$ disclination core with $s = -1$ for a magnetization pointing downwards and $s = +1$ for a magnetization pointing upwards.

\subsection{Skyrmion charge of helimagnetic domain walls}

After having clarified the topological charge of a single dislocation, we now turn to the discussion of skyrmion charges of various domain walls. In order to determine the total charge of a domain wall we can add the charges of dislocations contained within it. As we consider only periodic domain walls, we can limit ourselves to a discussion of the skyrmion charge associated with their primitive unit cell that we denote by $w$, see Fig.~S13 for various examples.

The skyrmion charge of curvature walls vanishes because these walls can be continuously transformed to plain helices by straightening out their lamellar pattern. Non-zero skyrmion charges can only appear for topologically non-trivial walls of type II and III. 

All type II walls contain a single $\pi$ and $-\pi$ disclination per primitive unit cell separated by a distance $D$. Such a 
pair of disclinations is topologically equivalent to a dislocation with Burgers vector $B = 2D$, see Fig.~1 of the main text. So we can conclude that the charge of type II walls per primitive unit cell is given by $w = \frac{s}{2} {\rm mod}_2 \frac{2 D}{\lambda}$ where $s = 1$ if the magnetization at the core of the $\pi$ disclination points up (bright) and $s = -1$ if it points down (dark).
As a result, we obtain an even/odd effect for type II walls as shown in Fig.~S13a-i representing such walls with decreasing values of $D$. For example in panel h the distance between the disclinations is a single helix wavelength, $D = \lambda$, so that $w = 0$, whereas in panel i it is only half a helix wavelength, $D = \lambda/2$, and $w = \frac{1}{2}$.
 
For the discussion of the skyrmion charge of type III dislocation walls we distinguish between situations where their lamellar stripes are broken or not. If they remain unbroken like in Fig.~S13j-m, which also includes the ground state configurations for large angles $\angle({\bf Q}_1, {\bf Q}_2)$, their total skyrmion charge always vanishes. In this case, the lamellar stripes, similar to curvature walls, can be straightened out and transformed to plain helices. The same result is obtained by adding the charges of dislocations within such type III walls. For example, in Fig.~S13j-l the dislocations both possess a Burgers vector that is an even integer multiple of $\lambda$ so that $w = 0$. In panel k and m, the Burgers vector is an odd integer multiple of $\lambda$ but the primitive unit cell contains two dislocations with opposite magnetizations at their cores (one bright and one dark). The finite charges of these two dislocations therefore compensate each other so that again $w = 0$.

A finite skyrmion charge, however, might be obtained for type III walls with broken stripes like in the examples shown in Fig.~S13n and o. They both possess two dislocations per primitive unit cell where each carries a charge $\frac{1}{2}$ resulting in $w = 1$.

\newpage

\section*{Supplementary Figures}

\begin{figure}
\centering
\includegraphics[width=0.4\textwidth]{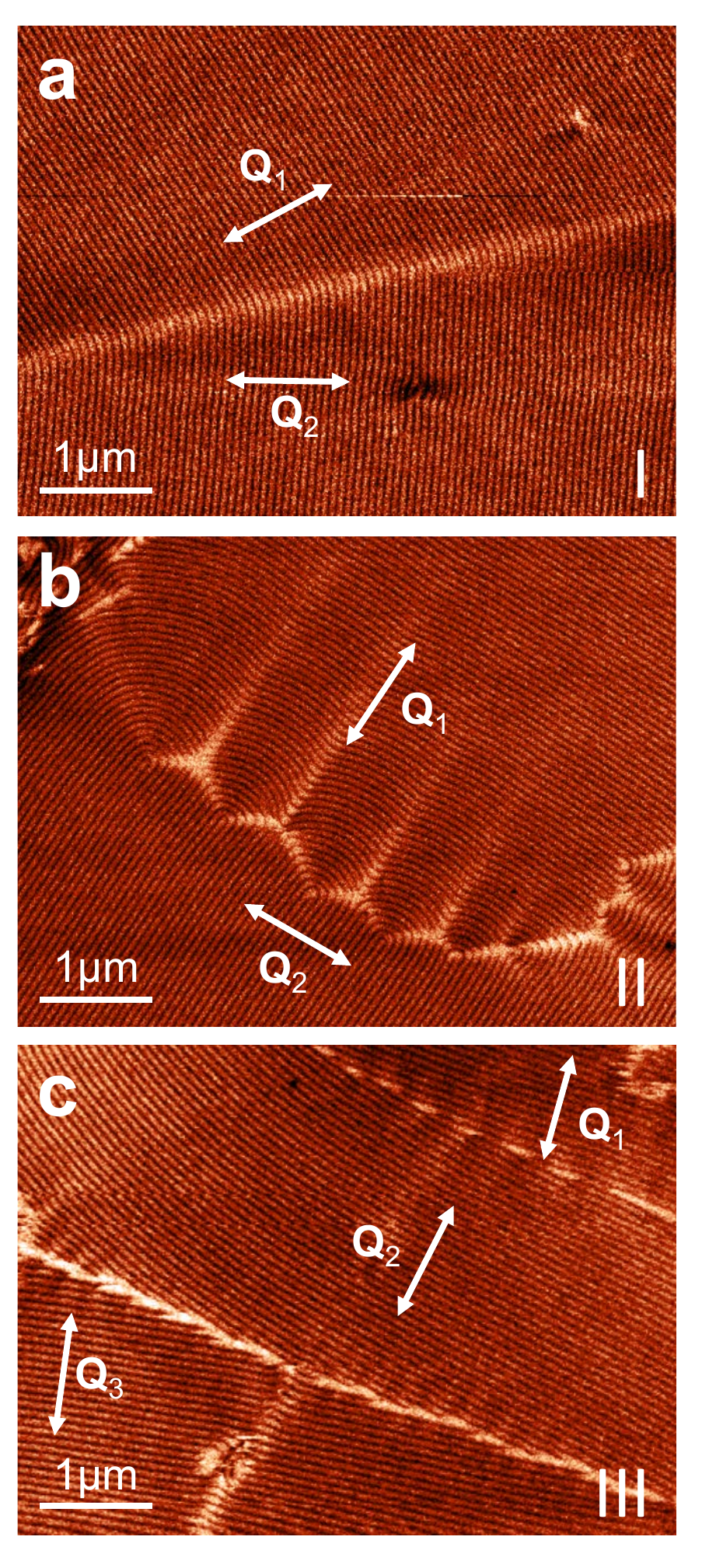}
\renewcommand\thefigure{S1}
\\
\caption{\label{fig:WallExamples}\\ {\bf Figure S1.}
{\bf MFM images of the three main domain wall configurations.} In comparison to type I and III, 
the width of domain walls of type II is relatively large comprising micrometer-sized distortions due to $-\pi$ disclinations.
}
\end{figure}
\newpage

\begin{figure}
\centering
\includegraphics[width=0.8\textwidth]{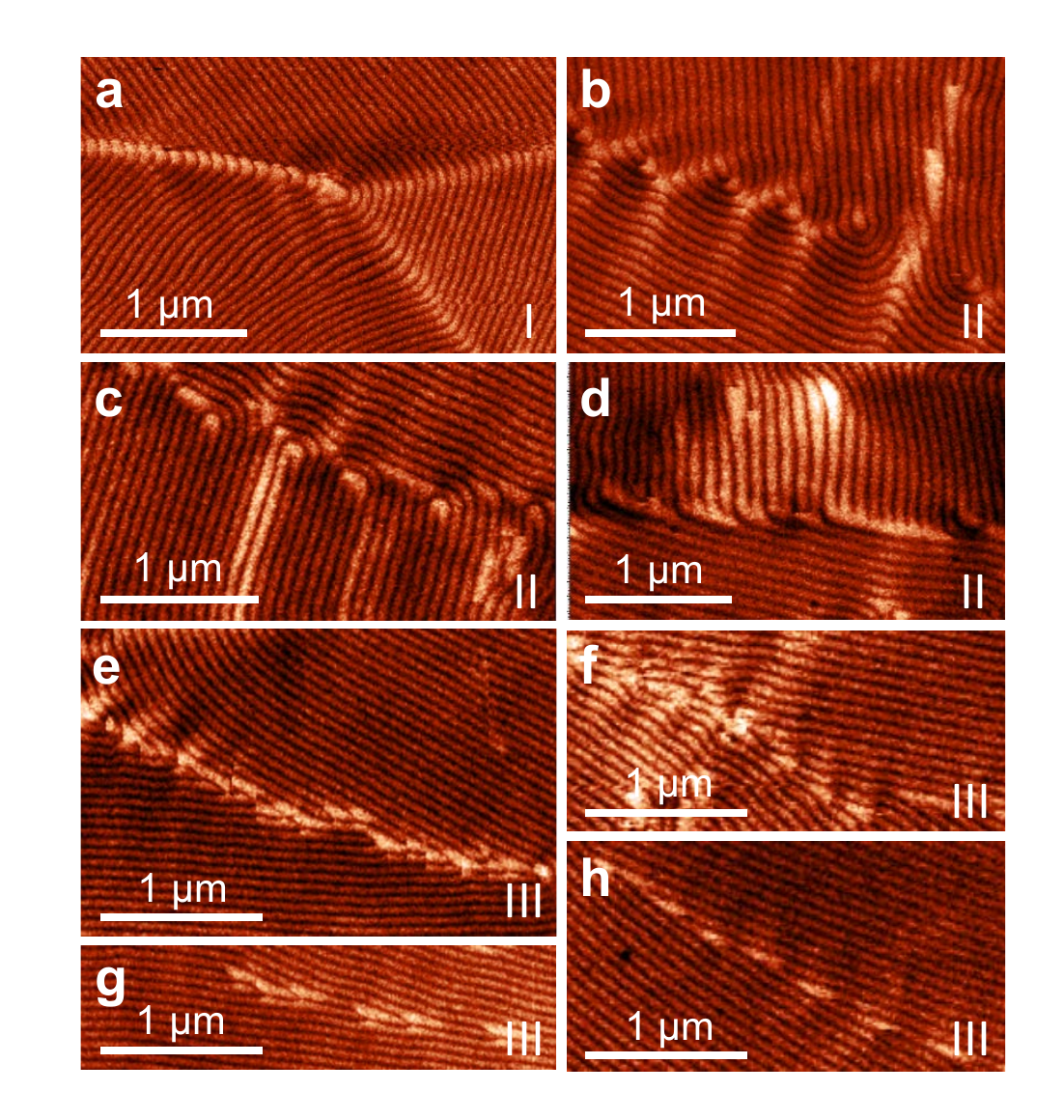}
\renewcommand\thefigure{S2}
\\
\caption{\label{fig:WallExamples2}\\ {\bf Figure S2.}
{\bf Domain wall configurations in FeGe.} The number of domain walls reflect the high variety of domain wall subtypes.
}
\end{figure}
\newpage
\begin{figure}
\centering
\includegraphics[width=0.6\textwidth]{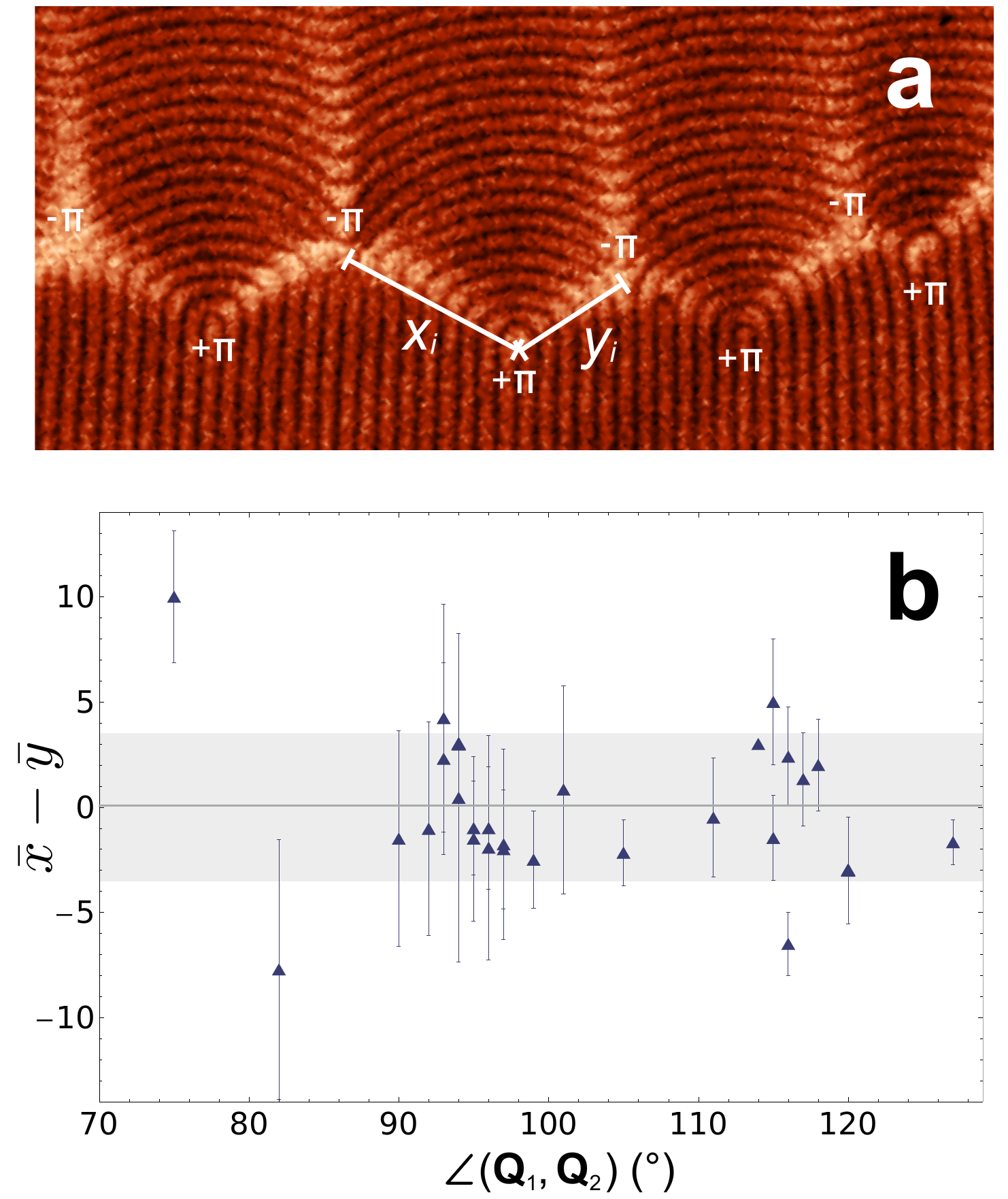}
\renewcommand\thefigure{S3}
\\
\caption{\label{fig:PiDistancesExp2}\\ {\bf Figure S3.} 
{\bf Inner structure of type II domain walls.} Panel {\bf a} displays a representative MFM image of a type II domain wall with a zig-zag chain of $\pm\pi$ disclinations. The distances between those defects are measured by $x$ and $y$ in units of $\lambda/2$. In panel {\bf b}, the difference between $x$ and $y$ averaged over each domain wall is shown as a function of $\angle({\bf Q}_1 ,{\bf Q}_2)$ for 27 experimentally observed type II walls. The grey area indicates the corresponding standard deviation.}
\end{figure}

\newpage

\begin{figure}
\centering
\includegraphics[width=0.8\textwidth]{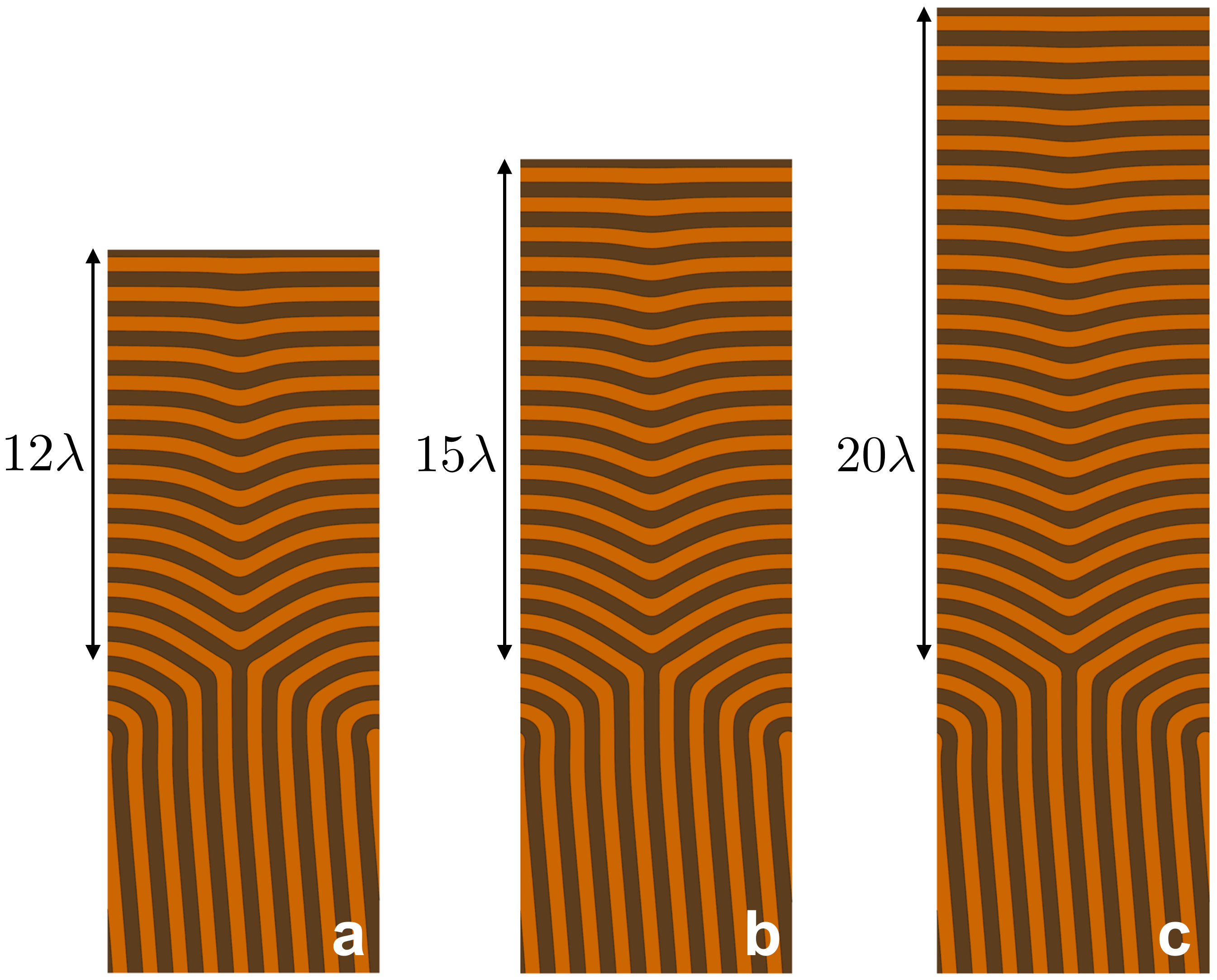}
\renewcommand\thefigure{S4}
\\
\caption{\label{fig:Num}\\ {\bf Figure S4.}
{\bf Micromagnetic simulations of a type II domain wall for different system sizes.} Optimized helimagnetic texture with a type II zig-zag disclination wall with $\angle({\bf Q}_1,{\bf Q}_2) = 94^\circ$ and a distance $D = 9\lambda/2$ between disclinations for three different system sizes with fixed boundary conditions at the lower and upper edge. A single period of the wall is shown where  bright and dark stripes represent regions where the magnetization points out or into the plane, respectively. For all three setups the number of sites along the wall was $N_x = 378$. The number of sites perpendicular to the wall is $N_{y} = 1008, 1134, 1344$ and the distance between the upper boundary and the domain wall interface is $L=12,15,20\lambda$, respectively, for panel {\bf a}, {\ bf b}, and {\bf c}.
}
\end{figure}

\newpage

\begin{figure}
\centering
\includegraphics[width=0.8\textwidth]{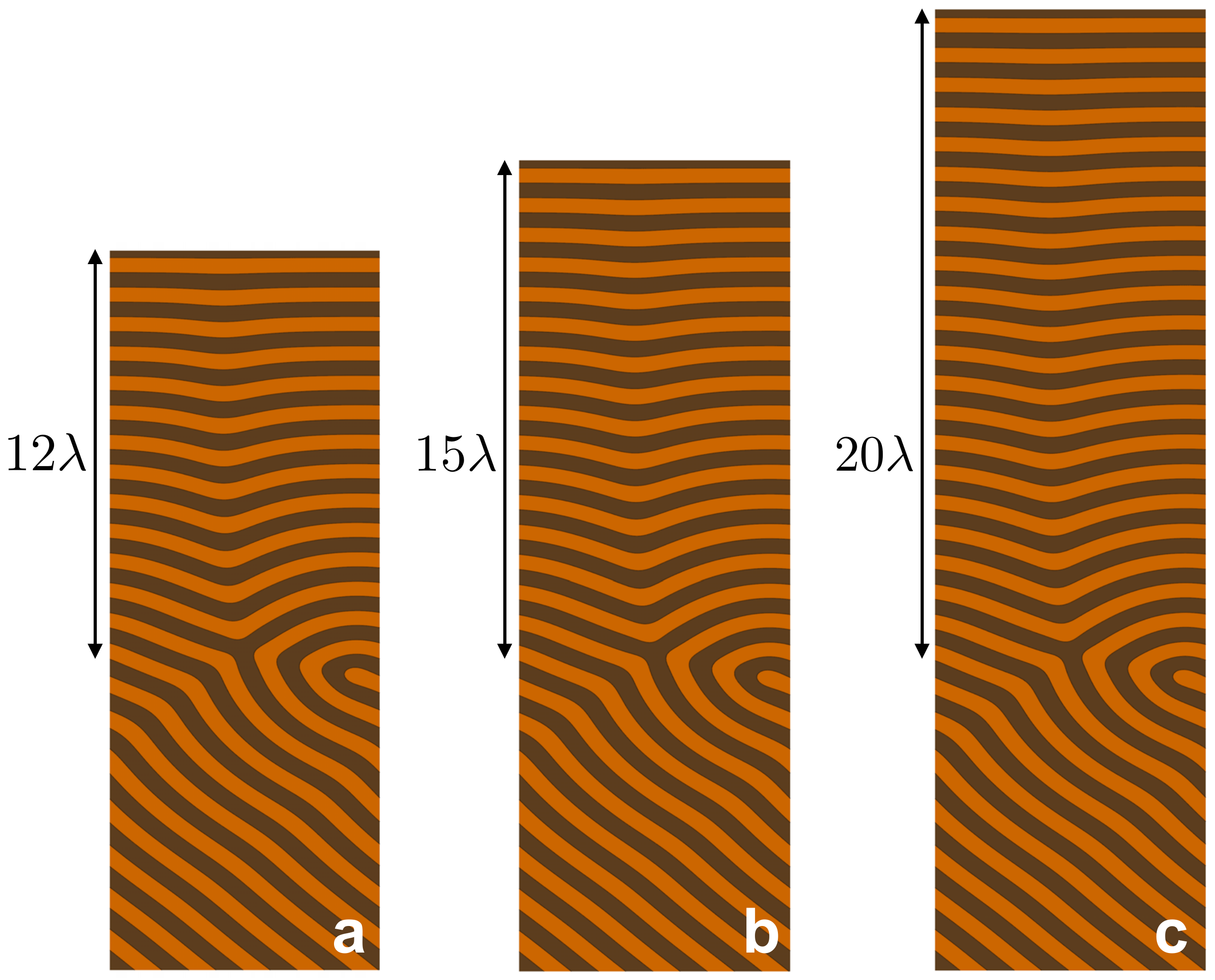}
\renewcommand\thefigure{S5}
\\
\caption{\label{fig:Num2}\\ {\bf Figure S5.}
{\bf Micromagnetic simulations of a type II domain wall with larger angle for different system sizes.} Optimized helimagnetic texture with a type II zig-zag disclination wall like in Fig.~S4 but for $\angle({\bf Q}_1, {\bf Q}_2) = 134^\circ$ and 
$D = 5\lambda/2$. The number of sites along the wall is $N_x = 295$, and perpendicular to the wall 
 they possess the same values as in Fig.~S4.
}
\end{figure}

\begin{figure}
\centering
\includegraphics[width=0.8\textwidth]{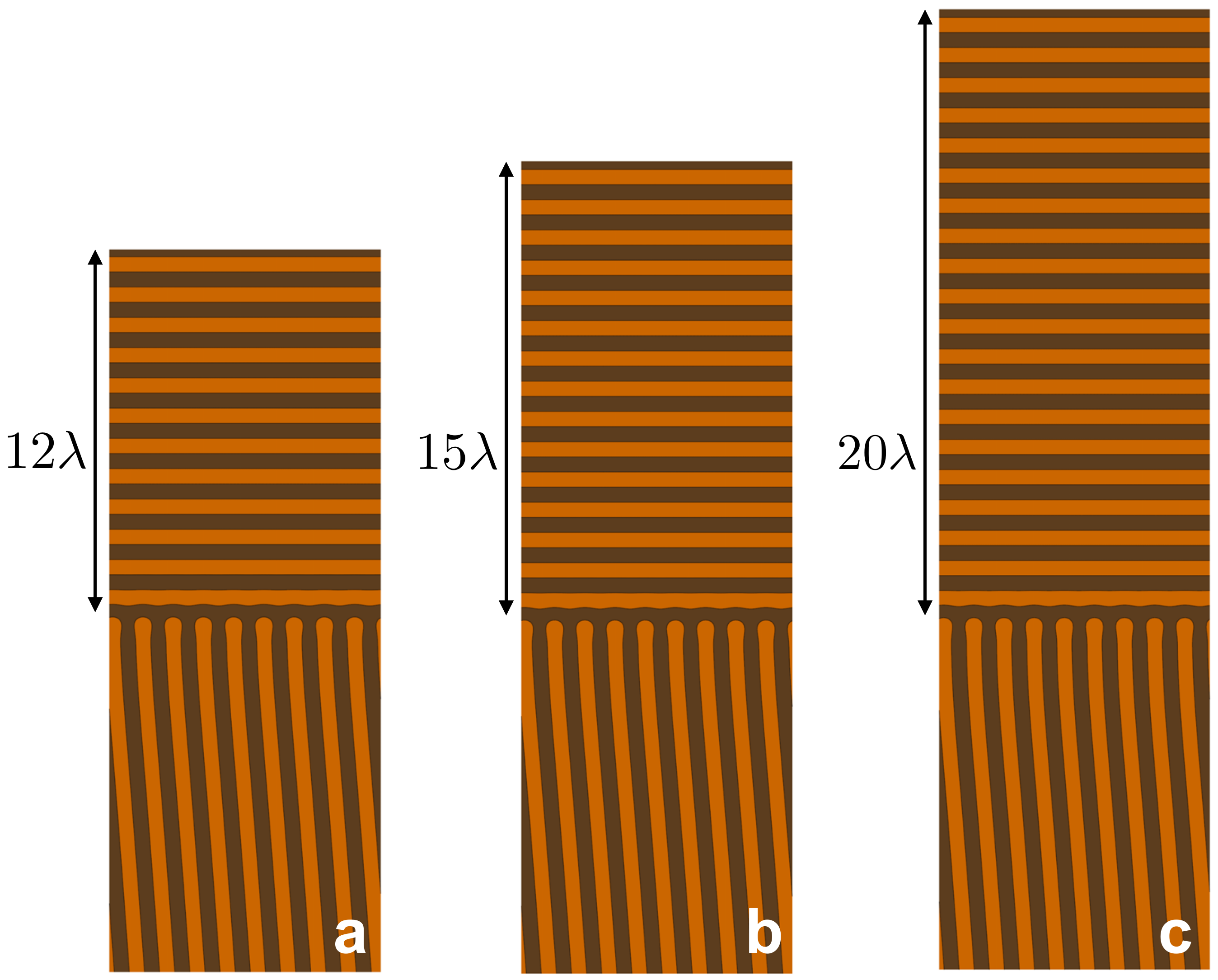}
\renewcommand\thefigure{S6}
\\
\caption{\label{fig:Num3}\\ {\bf Figure S6.}
{\bf Micromagnetic simulations of a type III domain wall for different system sizes.} Optimized helimagnetic texture with a type II zig-zag disclination wall like in Fig.~S4 with $\angle({\bf Q}_1, {\bf Q}_2)= 94^\circ$ and 
$D = \lambda/2$. The number of sites are the same as in Fig.~S4.
}
\end{figure}
\newpage
 \begin{figure}
\centering
\includegraphics[width=0.8\textwidth]{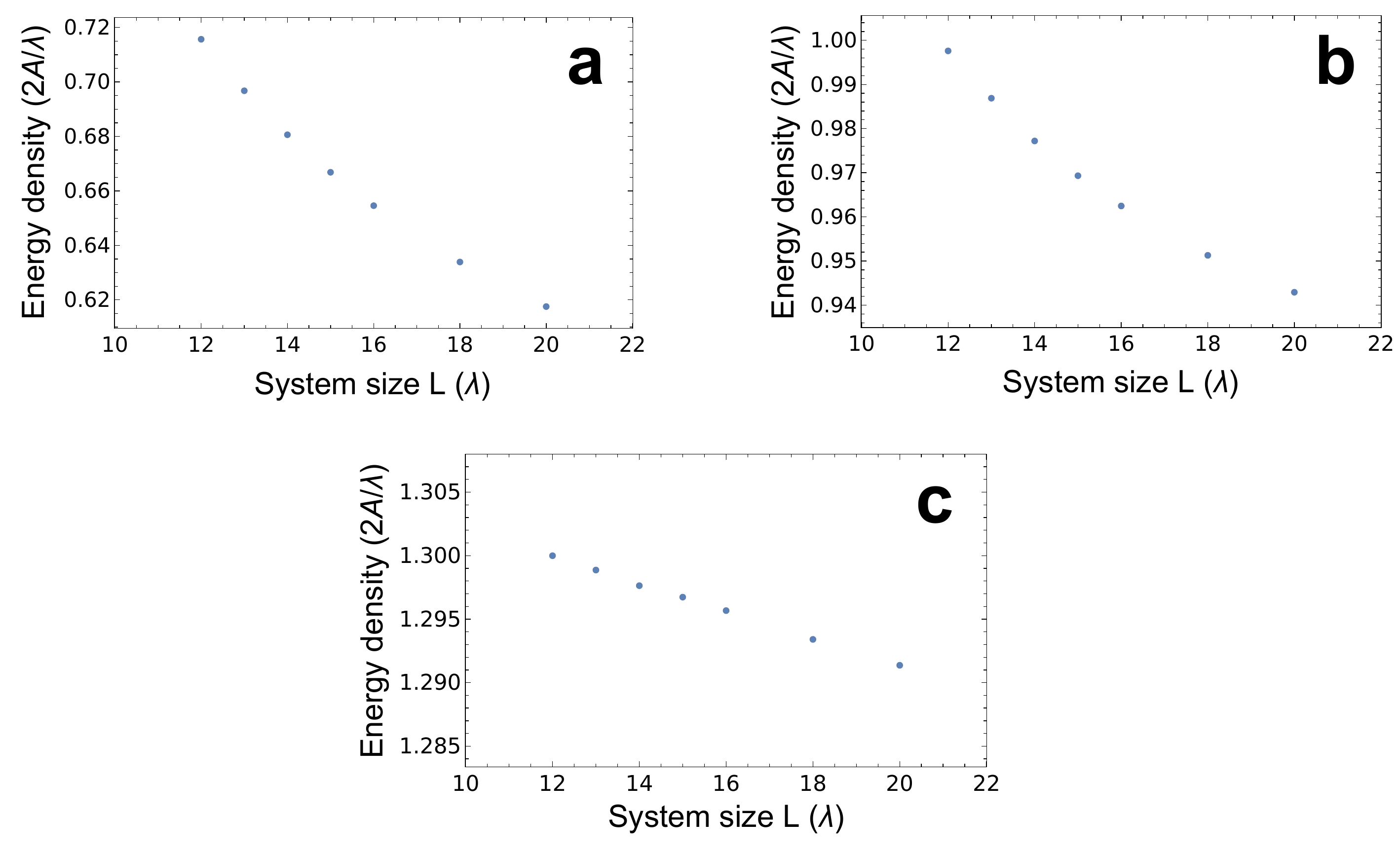}
\renewcommand\thefigure{S7}
\\
\caption{\label{fig:SizeScaling}\\ {\bf Figure S7.}
{\bf Finite size scaling of the domain-wall energy density.} Finite size scaling of the domain wall energy density for the three examples shown in Fig.~S4, Fig.~S5 and Fig.~S6 displayed in panel {\bf a}, {\bf b}, and {\bf c}, respectively.
}
\end{figure}

\newpage

\begin{figure}
\centering
\includegraphics[width=0.5\textwidth]{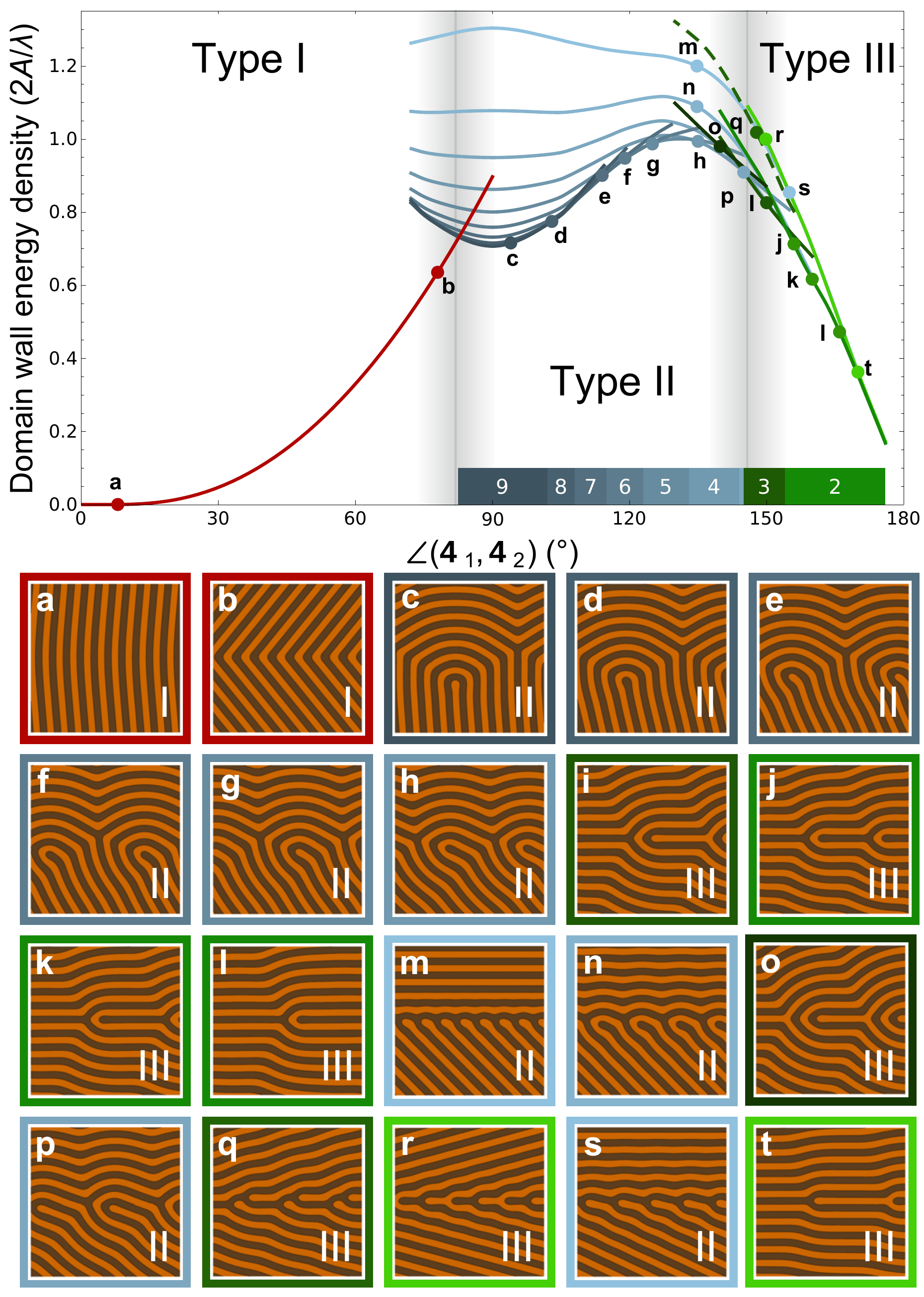}
\renewcommand\thefigure{S8}
\\
\caption{\label{fig:WallConf}\\ {\bf Figure S8.}
{\bf Domain-wall energy density.} Energy densities were obtained with the help of micromagnetic simulations for a system size $L = 12\lambda$ as shown in Fig. 3 of the main text. The blue and green shaded boxes above the $x$-axis refer to properties of domain walls with lowest energies for a given angle 
$\angle({\bf Q}_1, {\bf Q}_2)$. The blue shaded boxes indicate the separation $D$ of disclinations within type II walls in units of $\lambda/2$, and the green shaded boxes indicate the Burgers vector of serially aligned edge dislocations within type III walls in units of $\lambda$. Panels a to l represent a series of domain wall ground states as a function of increasing angle $\angle({\bf Q}_1, {\bf Q}_2)$. Panel m to t show domain wall configurations with higher energies. 
}
\end{figure}
\newpage
\begin{figure}
\centering
\includegraphics[width=0.8\textwidth]{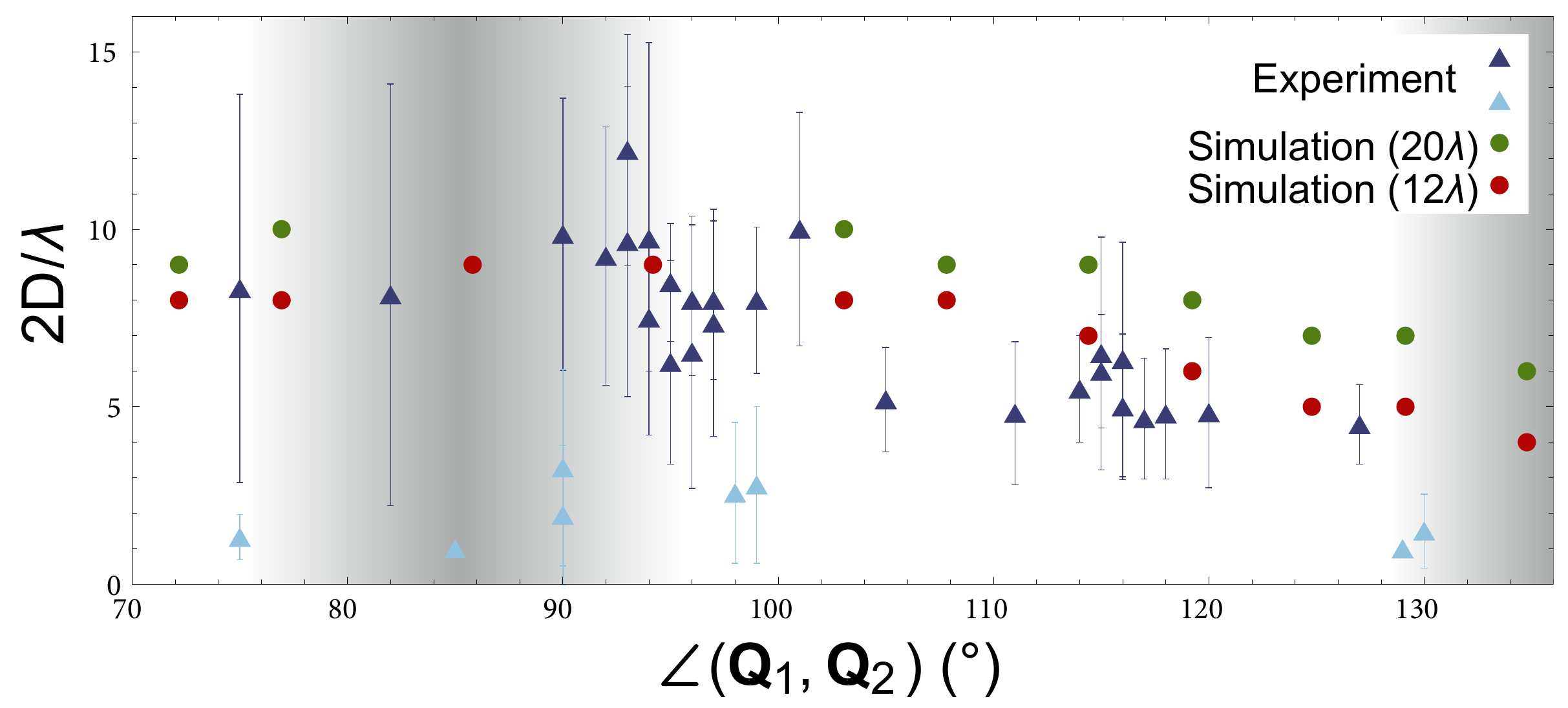}
\renewcommand\thefigure{S9}
\\
\caption{\label{fig:PiDistances} \\{\bf Figure S9.}
{\bf Experimental and theoretical distances between $\pi$ and $-\pi$ disclinations within type II domain walls.} Experimental values (light blue and blue triangles) are obtained in Section \ref{sec:StatAnalysis}. Theoretical values for special values of $\angle({\bf Q}_1, {\bf Q}_2)$ are shown for two implemented system sizes $L = 12\lambda$ (red circles) and $L = 20\lambda$ (green circles).}
\end{figure}

\newpage
\begin{figure}
\centering
\includegraphics[width=0.8\textwidth]{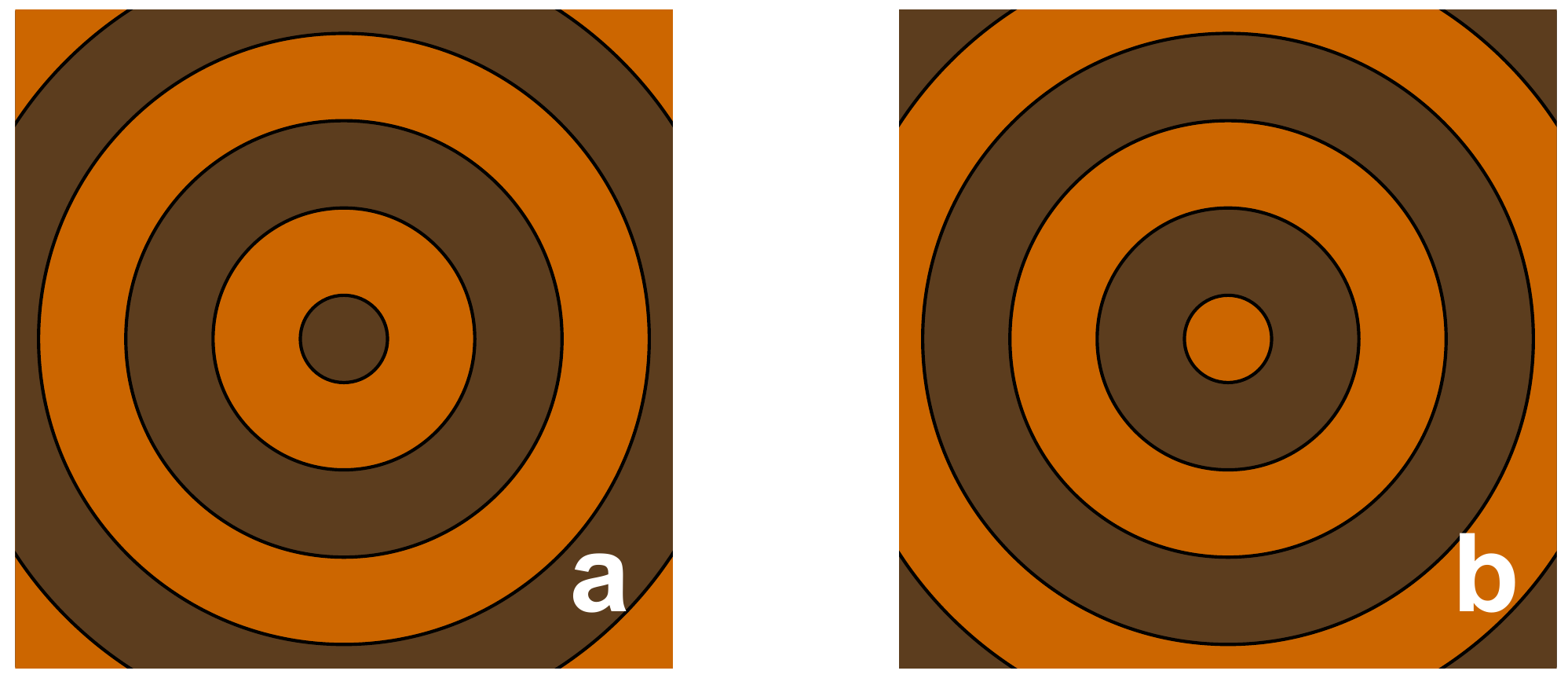}
\renewcommand\thefigure{S10}
\\
\caption{\label{fig:2piDisloc}\\ {\bf Figure S10.}
{\bf Illustrations of $2\pi$ disclinations.} $2\pi$ disclinations with ({\bf a}) a dark center corresponding to a magnetization pointing down ($\theta_0 = \pi$ in Eq.~\eqref{2pi}) and ({\bf b}) a bright center corresponding to a magnetization pointing up ($\theta_0 = 0$ in Eq.~\eqref{2pi}). These are only illustrations and not configurations optimimized by micromagnetics.}
\end{figure}

\begin{figure}
\centering
\includegraphics[width=0.8\textwidth]{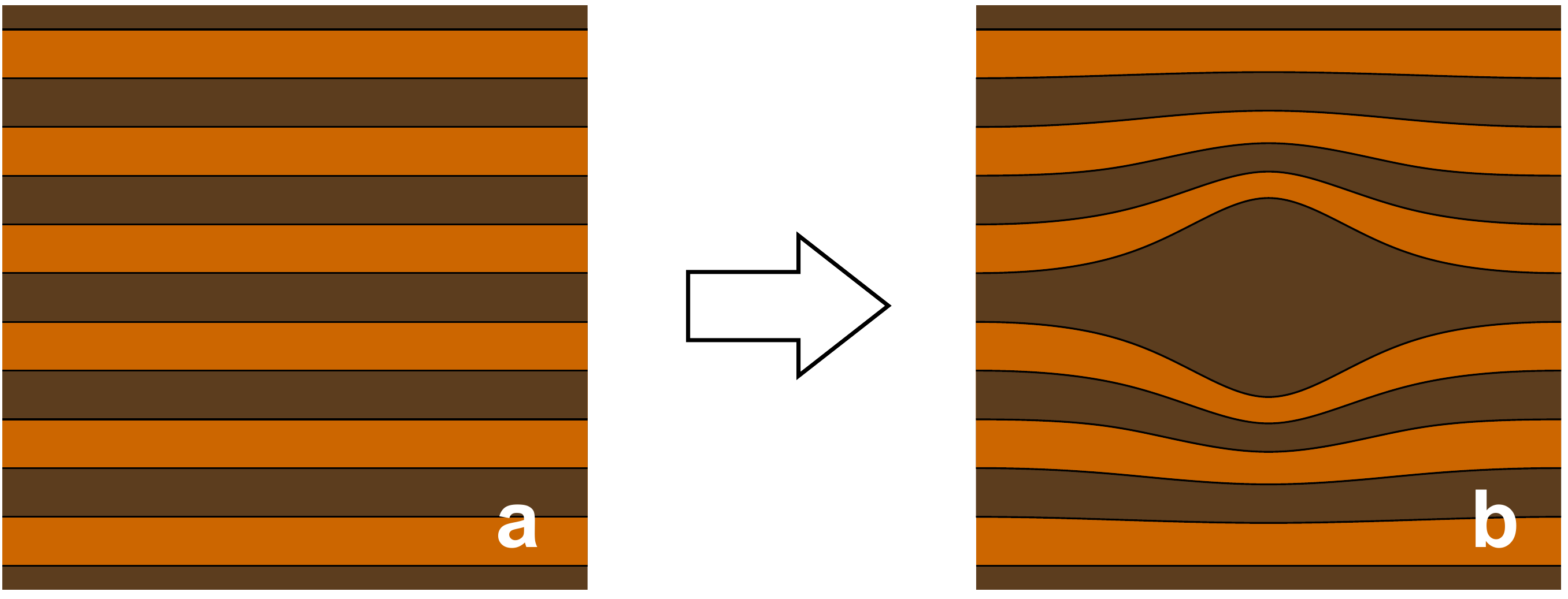}
\renewcommand\thefigure{S11}
\\
\caption{\label{fig:Transform}\\ {\bf Figure S11.}
{\bf Elastic deformation of the helimagnetic structure.} The pristine helix ({\bf a}) and the elastically deformed configuration ({\bf b}) both possess vanishing total skyrmion charge. These are only illustrations and not configurations optimized by micromagnetics.}
\end{figure}

\newpage

\begin{figure}
\centering
\includegraphics[width=0.8\textwidth]{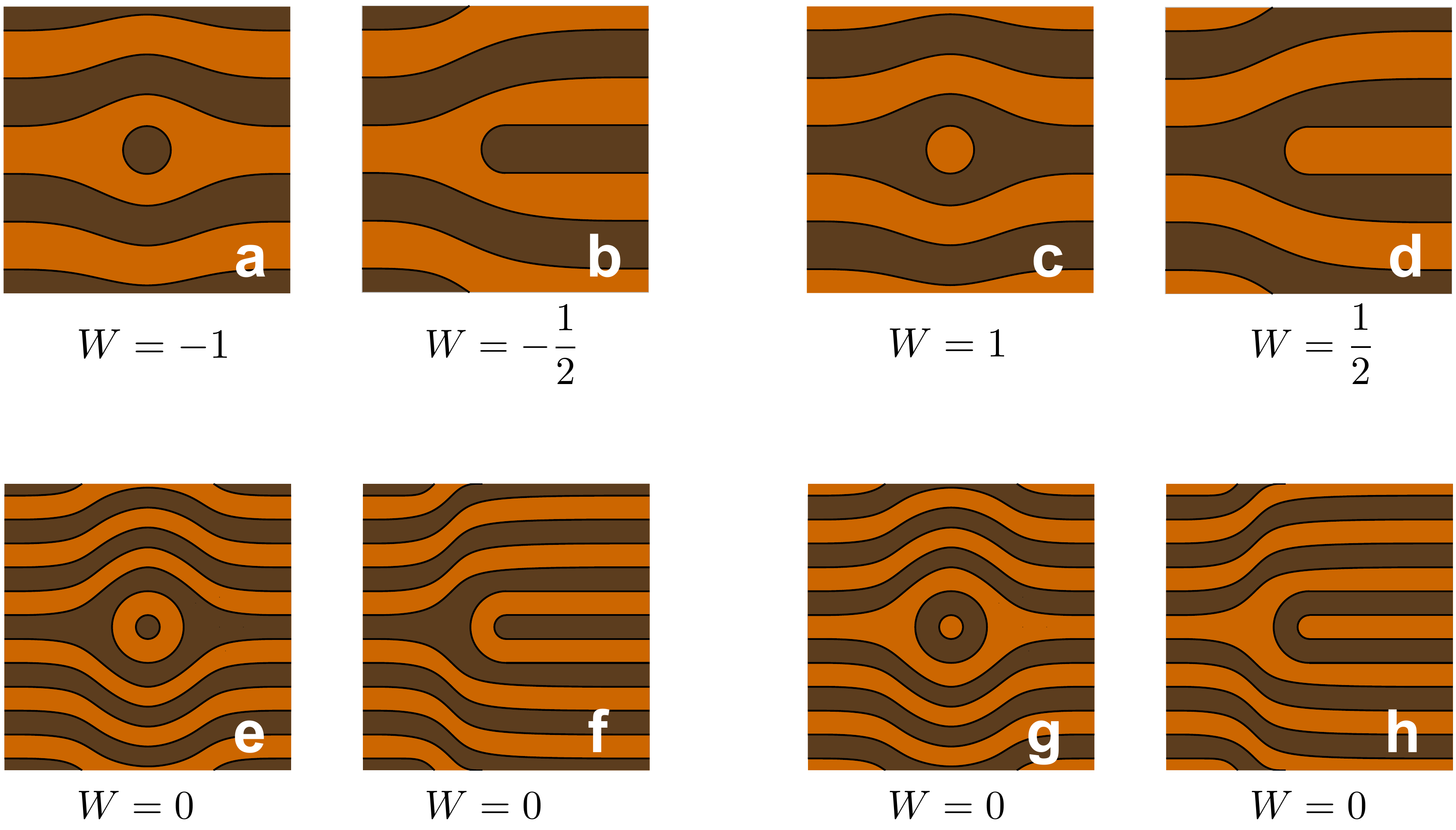}
\renewcommand\thefigure{S12}
\\
\caption{\label{fig:DislocExamples}\\ {\bf Figure S12.}
{\bf Visual derivation of the total skyrmion charge for a dislocation.} A $2\pi$ disclination with finite radius $m \pi/Q = m \lambda/2$ embedded in a magnetic helix structure can be interpreted as two dislocations with Burgers vector $B = m \lambda$ each carrying a charge $W$ defined in Eq.~\eqref{ChargeDisl}. {\bf a} and {\bf c} show an embedded $2\pi$ disclination with radius $\lambda/2$, i.e., a skyrmion where the magnetization at the core is pointing downwards (dark) and upwards (bright) resulting in a total skyrmion charge $W = -1$ and $W = 1$, respectively. The corresponding dislocations in {\bf b} and {\bf d} each possess half this value of the topological charge. In contrast, the total skyrmion charge of an embedded $2\pi$ disclination with radius $\lambda$ and of its associated dislocations vanishes, see panel {\bf e}-{\bf h}. These are only illustrations and not configurations optimized by micromagnetics.
}
\end{figure}
\newpage
\begin{figure}
\centering
\includegraphics[width=0.8\textwidth]{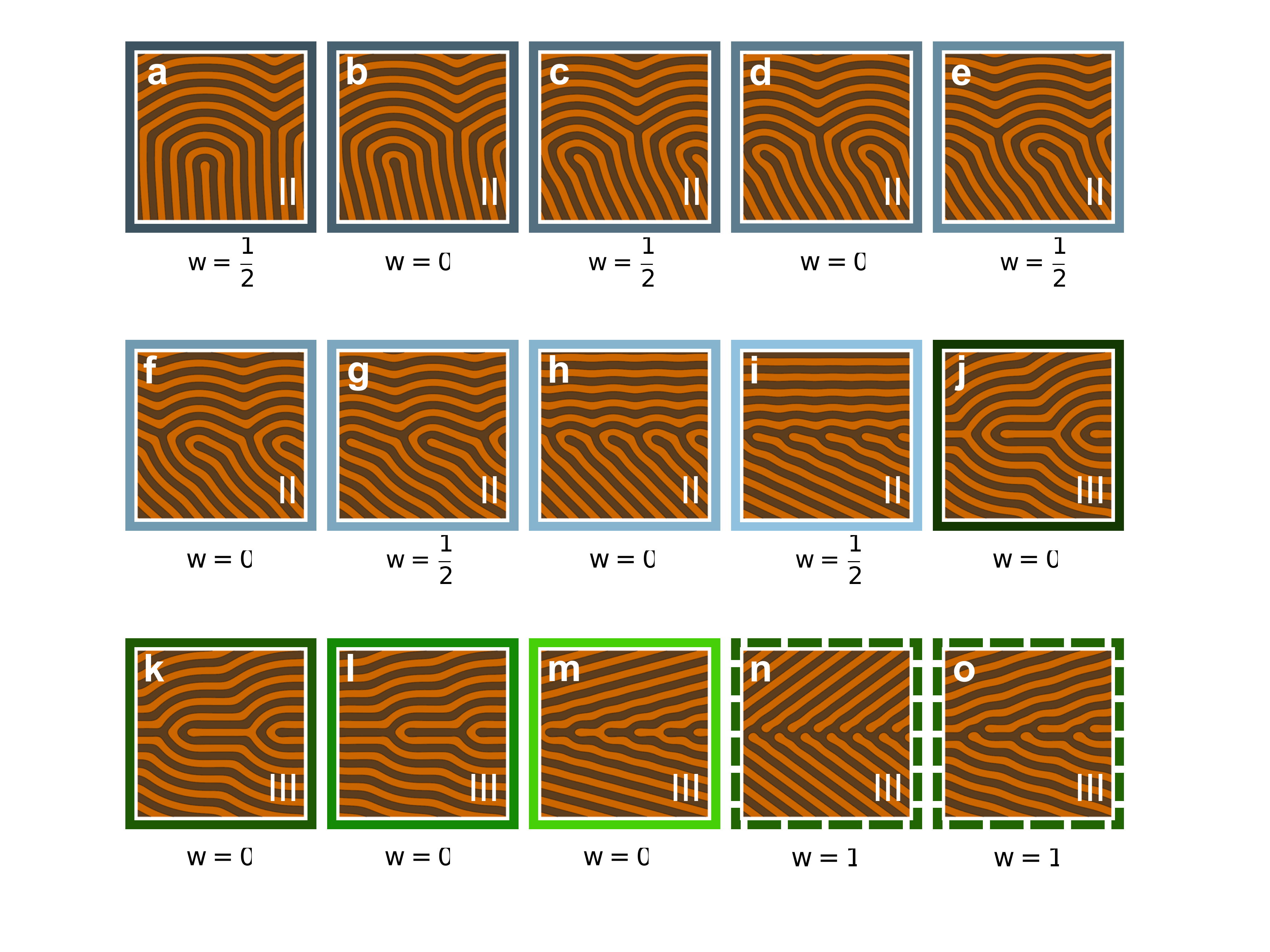}
\renewcommand\thefigure{S13}
\\
\caption{\label{fig:WallsCharge}\\ {\bf Figure S13.}
{\bf Skyrmion charge for all domain wall types.} Examples of periodic domain walls and total skyrmion charges $w$ associated with their primitive unit cell. Note that the panels display a region of the walls that is larger than the extent of a single primitive unit cell.}
\end{figure}

\end{document}